\documentclass[a4paper, 12pt, oneside]{article}

\usepackage{jheppub}

\usepackage{graphicx,wrapfig,float,slashed,subcaption,bbold,bm}
\usepackage{amsmath,amssymb,epsfig,graphicx,xcolor}
\usepackage[normalem]{ulem}
\usepackage{epstopdf}
\usepackage{booktabs}
\usepackage{ragged2e}
\usepackage{mciteplus}
\usepackage{mathtools}
\usepackage{float}
\allowdisplaybreaks

\usepackage{cancel}

\usepackage{makecell}
\usepackage{multirow}
\usepackage{adjustbox}

\newcommand{\nc}{\newcommand}
\nc{\non}{\nonumber}
\nc{\hc}{\hbox {H.c.}}
\nc{\noi}{\noindent}
\nc{\barx}{\bar{x}}
\nc{\pbarn}{\;\hbox {pb}}
\nc{\fbarn}{\;\hbox {fb}}

\nc{\hsp}{\hspace{0.5cm}}
\nc{\lsp}{\hspace{1cm}}
\nc{\Lsp}{\hspace{2cm}}
\nc{\LLsp}{\lsp\lsp}
\nc{\lra}{\longrightarrow}
\nc{\p}{\prime}
\nc{\sgn}{\text{sgn}}
\nc{\ph}{\varphi}
\nc{\op}{{\cal O}}

\nc{\beq}{\begin{equation}}  
\nc{\eeq}{\end{equation}}
\nc{\bea}{\begin{eqnarray}}  
\nc{\eea}{\end{eqnarray}}
\nc{\baa}{\begin{array}}     \nc{\eaa}{\end{array}}
\nc{\bit}{\begin{itemize}}   \nc{\eit}{\end{itemize}}
\nc{\ben}{\begin{enumerate}} \nc{\een}{\end{enumerate}}
\nc{\bce}{\begin{center}}    \nc{\ece}{\end{center}}
\nc{\bpm}{\begin{pmatrix}}   \nc{\epm}{\end{pmatrix}}

\def\lsim{\mathrel{\raise.3ex\hbox{$<$\kern-.75em\lower1ex\hbox{$\sim$}}}}
\def\gsim{\mathrel{\raise.3ex\hbox{$>$\kern-.75em\lower1ex\hbox{$\sim$}}}}

\def\udots{\mathinner{\mkern1mu\raise1pt\vbox{\kern7pt\hbox{.}}\mkern2mu\raise4pt\hbox{.}\mkern2mu\raise7pt\hbox{.}\mkern1mu}}

\newcommand{\Eq}[1]{Eq.~(\ref{#1})}

\newcommand{\bealg}{\begin{equation}\begin{aligned}} 
\newcommand{\eealg}{\end{aligned}\end{equation}}

\allowdisplaybreaks

\preprint{\begin{flushright}
UT-WI-32-2026\\
\end{flushright}}

\title{Constraints on Decaying Dark Matter with $B=1$ from Fermi-LAT}
\author[a]{Can Kilic,}
\author[a]{Zhi-Xiang Yang,}
\author[b,c]{and Taewook Youn}
\affiliation[a]{Theory Group, Weinberg Institute for Theoretical Physics, University of Texas at Austin, Austin, TX 78712, U.S.A.}
\affiliation[b]{Gravity, Spacetime, and Particle Physics (GRASP) Initiative \\
Harvard University, 17 Oxford Street, Cambridge, MA 02138, USA}
\affiliation[c]{School of Physics, Korea Institute for Advanced Study \\  Seoul 02455, Republic of Korea}
\emailAdd{kilic@physics.utexas.edu}
\emailAdd{zxiang@utexas.edu}
\emailAdd{taewookyoun@fas.harvard.edu}

\abstract{
We investigate the gamma ray signatures of fermionic dark matter that carries one unit of baryon number and decays into three quarks. We compute the expected prompt and secondary gamma-ray spectra for a representative range of final state flavor combinations. By comparing these spectra to Fermi-LAT measurements of the isotropic gamma-ray background, we derive exclusion limits on the lifetime. We observe that depending on the dark matter mass and the flavor of the final state, the lifetime bounds on decays of $B=1$ dark matter can differ by up to a factor of two from those on decays of $B=0$ dark matter. Furthermore, light-quark final states generally yield stronger bounds at lower masses, whereas heavy-quark final states can yield comparable or stronger bounds at masses of tens of TeV and above.} 

\begin{document}

\maketitle
\flushbottom
\section{Introduction}
\label{Introduction}

The fundamental nature of dark matter (DM) remains one of the most significant unresolved questions in modern physics. While it is commonly taken for granted that the DM particle is exactly stable, it is worth remembering that the only empirical statement we can make is that it is very long lived, and very weak interactions of the DM particle may cause it to decay at very late times~\cite{Doroshkevich:1984}. If DM does decay, not merely within a decoupled dark sector but to a final state including Standard Model (SM) particles, this would result in observable signatures in cosmic-ray spectra. The scenario of decaying dark matter (DDM) is consistent with bounds from direct-detection and collider searches~\cite{Liu:2016ngs, Moulin:2013saa, Blanco:2018esa, Cohen:2016uyg, Fornasa:2015qua}.

While the simplest models where DM is charged under the SM electroweak gauge interaction are disfavored~\cite{Cirelli:2005uq,Cirelli:2007xd,Cirelli:2009uv,Garcia-Cely:2015dda,Fan:2013faa,Cirelli:2015bda}, it is an intriguing possibility that DM may carry a charge under a global symmetry of the SM such as baryon number. If the DM particle is a fermion with $B=1$ and a mass above the GeV scale, it can decay into three quarks, which would subsequently hadronize. Remaining agnostic to the nature of the UV physics that leads to this decay channel, the phenomenology of this scenario can be studied in generality by including a dimension-six effective interaction term in the low energy Lagrangian such as~\cite{Davoudiasl:2010am, Zhao:2014nsa, Fornal:2020poq, Farina:2016ndq, Kilic:2021zqu, Alonso-Alvarez:2021oaj}: 
\begin{equation}
    \mathcal{L} \supset \frac{\lambda_{ijk}}{\Lambda^2} \left(\chi d^{c}_i\right) \left(u^{c}_j d^{c}_k\right) + \text{h.c.}\,,
    \label{eq:operator}
\end{equation}
where $\Lambda$ represents the scale of the UV physics, the brackets denote Lorentz singlets, $\chi$ is the dark matter particle (a left-handed Weyl spinor), $u^{c}_i$ and $d^{c}_j$ (also left-handed Weyl spinors) represent the $SU(2)$ singlet up-type and down-type quark fields (with flavor indices $i, j, k$), and $\lambda_{ijk}$ are dimensionless couplings. We use the two-component spinor notation of Ref.~\cite{Dreiner:2008tw}.

This interaction allows for eighteen distinct flavor combinations of quarks in the final state. 

As the quarks from the DM decay hadronize, and as those hadrons cascade decay down to stable particles, a continuous flux of protons, electrons, positrons, neutrinos, and gamma rays are created. Gamma rays provide a particularly powerful probe for indirect detection because they propagate in a straight line, without being affected by galactic magnetic fields.
The Fermi Large Area Telescope (Fermi-LAT) collaboration has provided precision measurements of the isotropic gamma-ray background (IGRB)~\cite{Fermi-LAT:2014ryh}, which represents the residual diffuse emission remaining after the subtraction of resolved point sources and modeled Galactic foregrounds. 
The IGRB serves as an optimal dataset for placing stringent constraints on the lifetime of decaying dark matter.

In this work, we systematically study the gamma-ray signatures of DDM with $B=1$ kinematically accessible over a mass range from 10 GeV up to 100 TeV. We calculate the expected prompt and secondary gamma-ray fluxes for all eighteen allowed flavor combinations and compare them against Fermi-LAT IGRB measurements. To provide a comprehensive evaluation, we derive $3\sigma$ exclusion limits on the DM lifetime using three distinct statistical frameworks: a conservative background independent approach, a background model with a combined power-law and exponential spectrum, and a likelihood analysis combining several models of distinct background components. While the lifetime bounds on $B=1$ dark matter are comparable to those on $B=0$ dark matter, we find that across the range of DM mass and final states we consider, the two bounds may differ by up to a factor of two. In addition, we find that light-quark final states generally yield stronger bounds at lower masses, whereas heavy-quark final states can yield comparable or stronger bounds at masses of tens of TeV and above, as the Fermi-LAT energy range probes progressively lower $x=E/M$.

Indirect detection searches for DDM have a long history, and the gamma-ray channel in particular has been used extensively to constrain the DM lifetime across a wide range of final states (see Ref.~\cite{Ibarra:2013cra} for a review, and Refs.~\cite{Cohen:2016uyg, Blanco:2018esa, Cholis:2024hmd} for more recent analyses). Most of this effort has targeted two-body decays into a pair of SM particles. Hadronic three-body decays of the type relevant here have received comparatively less attention: Ref.~\cite{Garny:2012vt} derived lifetime limits for three-body decays into quark final states, finding that the bounds depend on the final-state flavor and strengthen mildly for heavier quarks. Three-body decays have also been studied as a means of softening cosmic-ray spectra in the context of the AMS-02 positron excess~\cite{Cheng:2016slx}. Our work extends this program by performing a systematic, flavor-resolved analysis of a representative range of $B=1$ three-quark final states against the Fermi-LAT IGRB, using three complementary statistical frameworks.

This paper is organized as follows. In Section~\ref{Gamma Rays From Dark Matter Decay}, we describe our modeling of the signal and we compute the expected prompt and secondary gamma-ray fluxes originating from both galactic and extragalactic environments. In Section~\ref{Isotropic Gamma-Ray Background and Statistical Methods}, we describe the three statistical methodologies that we use for limit setting, and we present and discuss the resulting exclusion limits on the dark matter lifetime for all flavor combinations. We conclude in Section~\ref{Conclusion}.

\section{Modeling the Gamma Ray Flux from Dark Matter Decay}
\label{Gamma Rays From Dark Matter Decay}

To evaluate the observable signatures of DDM carrying baryon number, we must first determine the expected gamma-ray flux. In this section, we present the prompt spectra for each final state and we subsequently simulate both the galactic and extragalactic gamma-ray fluxes. When a DM particle decays into three final-state quarks, these quarks hadronize, and the hadrons may decay in multiple stages until only stable particles remain, including gamma rays, electrons, and positrons. The total observable gamma-ray signal has two components. The first contains prompt gamma rays that are produced directly during the hadronization and decay of the final-state quarks. In the second, the prompt electrons and positrons upscatter ambient background photons, such as the cosmic microwave background (CMB), starlight, and dust infrared emission, to gamma-ray energies via inverse Compton scattering (ICS). We simulate these prompt and secondary emissions originating from both the Galactic halo and extragalactic regions outside the Milky Way.

\subsection{Production}
\label{Production}

We implement the interaction of \Eq{eq:operator} in FeynRules~\cite{Alloul:2013bka} by integrating in a heavy color-triplet, electroweak-singlet scalar mediator $\phi$ with $(\mathbf{3},\mathbf{1}, -1/3)$, from which we generate a UFO (Universal FeynRules Output) model in \texttt{MG5\_aMC@NLO}~\cite{Degrande:2011ua, Alwall:2014hca}. For each DM mass and flavor channel, we generate $5\times10^4$ unweighted parton-level dark-matter decay events.

These events are passed to \textsc{Pythia8}~\cite{Bierlich:2022pfr}
for parton showering, hadronization, and the remaining unstable-particle decays, with the electroweak shower enabled.\footnote{For channels containing top quarks, MG5 generates the top and $W$ decays with spin correlations.} We construct the spectra used in the flux calculation by histogramming the stable final-state photons, electrons, positrons, and protons.

\begin{figure}[!htbp]
\centering

\begin{subfigure}{0.32\textwidth}
    \centering
    \includegraphics[width=\linewidth]{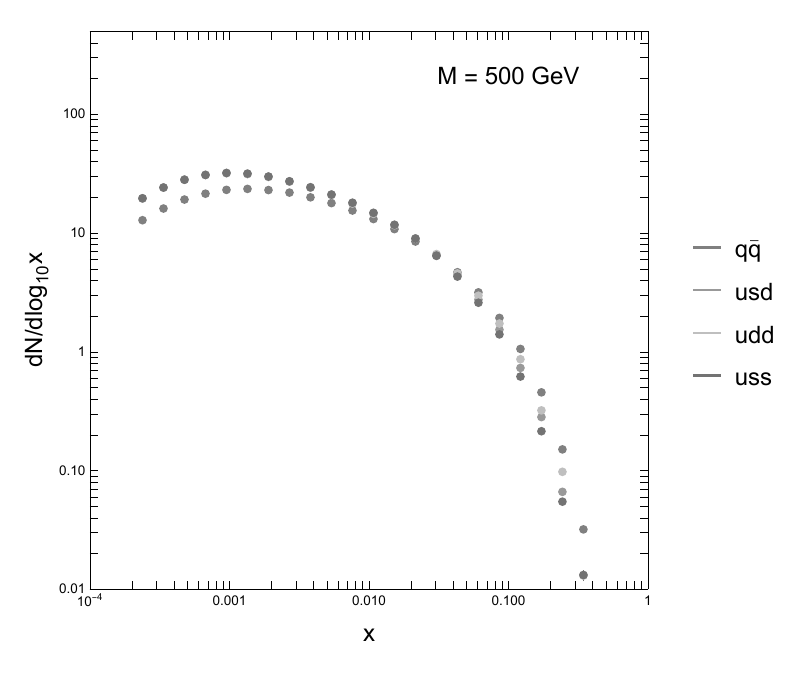}
\end{subfigure}
\hspace*{-1.18cm}
\begin{subfigure}{0.32\textwidth}
    \centering
    \includegraphics[width=\linewidth]{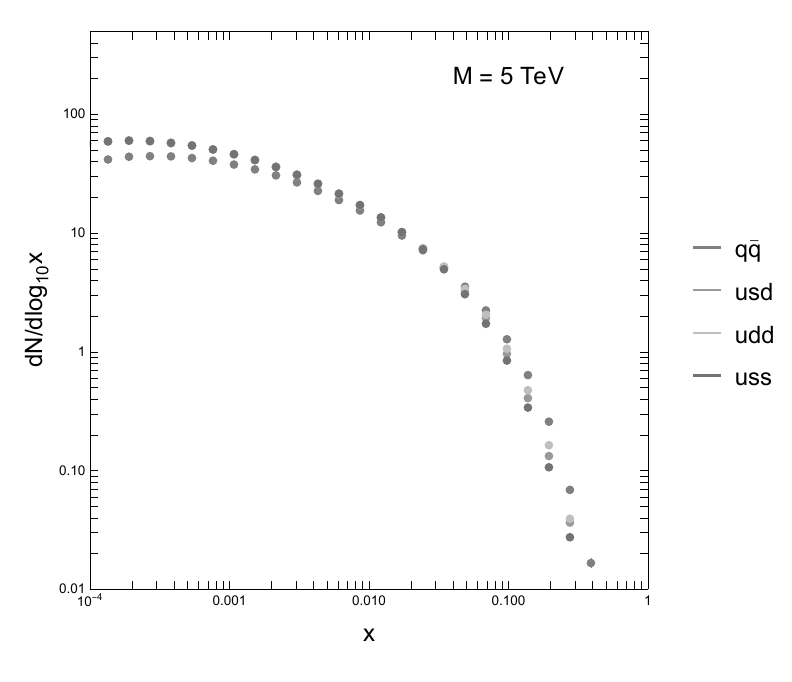}
\end{subfigure}
\hspace*{-1.18cm}
\begin{subfigure}{0.32\textwidth}
    \centering
    \includegraphics[width=\linewidth]{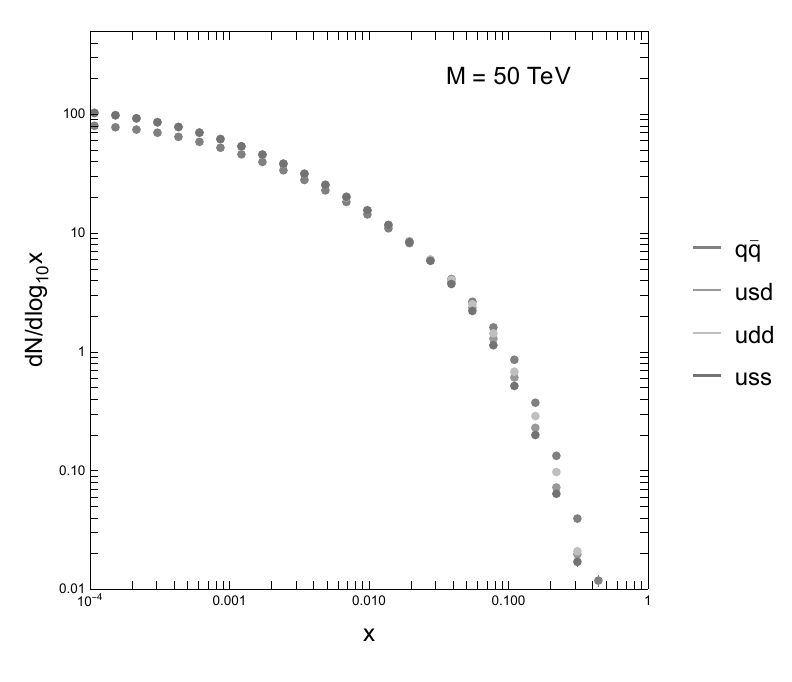}
\end{subfigure}

\vspace{0.5em}

\begin{subfigure}{0.32\textwidth}
    \centering
    \includegraphics[width=\linewidth]{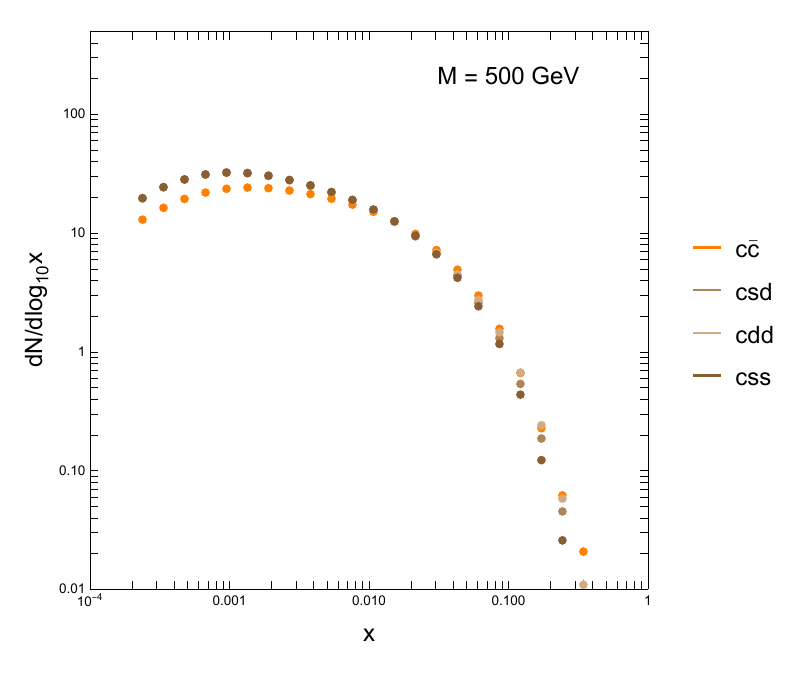}
\end{subfigure}
\hspace*{-1.18cm}
\begin{subfigure}{0.32\textwidth}
    \centering
    \includegraphics[width=\linewidth]{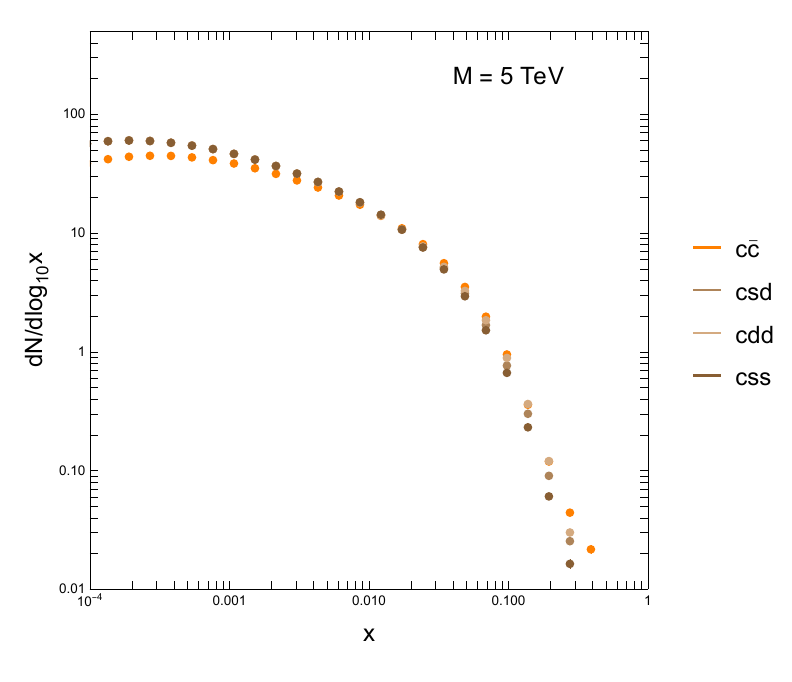}
\end{subfigure}
\hspace*{-1.18cm}
\begin{subfigure}{0.32\textwidth}
    \centering
    \includegraphics[width=\linewidth]{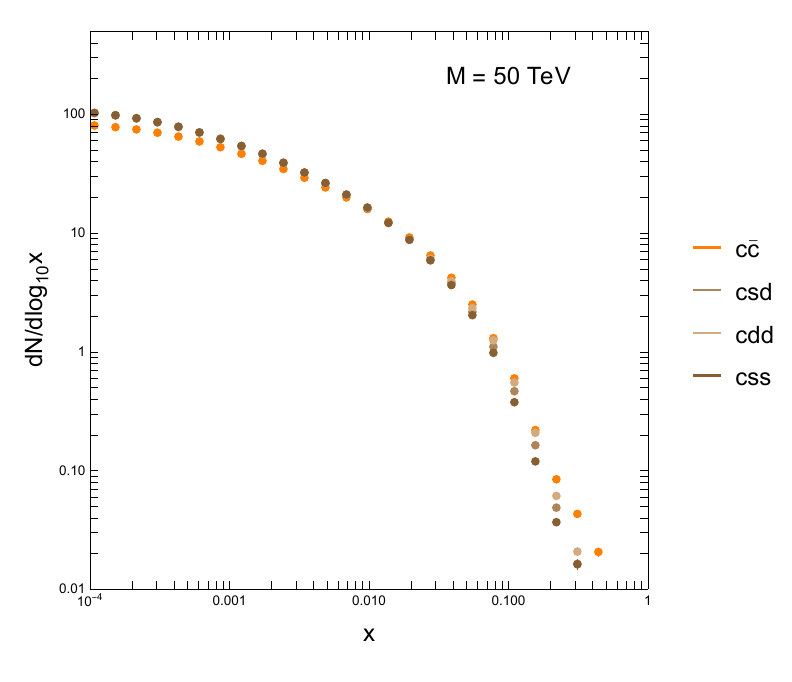}
\end{subfigure}

\vspace{0.5em}

\begin{subfigure}{0.32\textwidth}
    \centering
    \includegraphics[width=\linewidth]{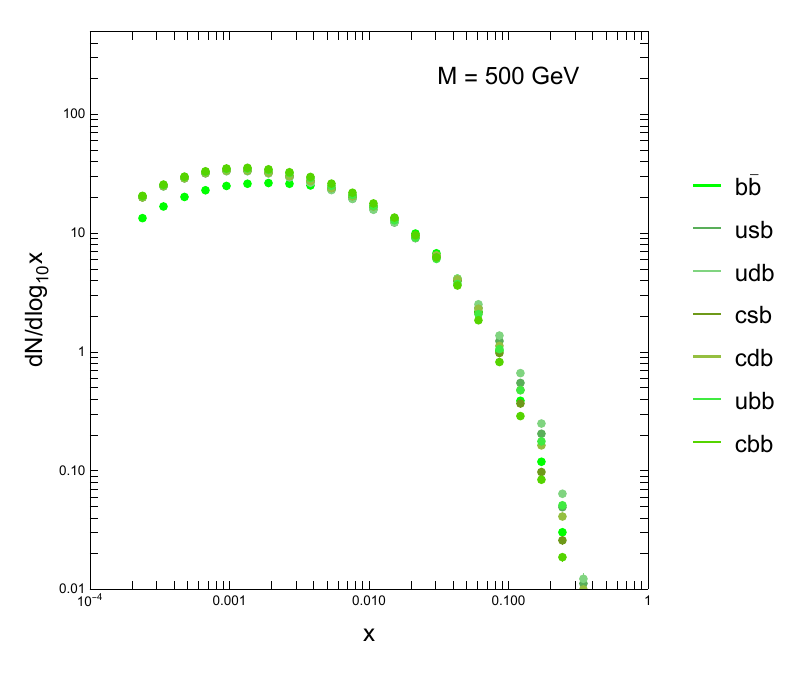}
\end{subfigure}
\hspace*{-1.18cm}
\begin{subfigure}{0.32\textwidth}
    \centering
    \includegraphics[width=\linewidth]{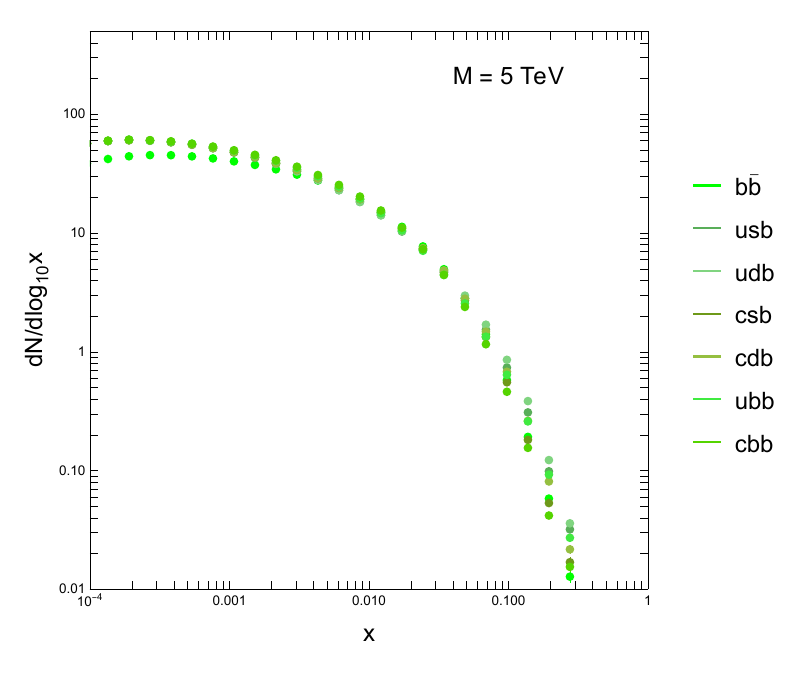}
\end{subfigure}
\hspace*{-1.18cm}
\begin{subfigure}{0.32\textwidth}
    \centering
    \includegraphics[width=\linewidth]{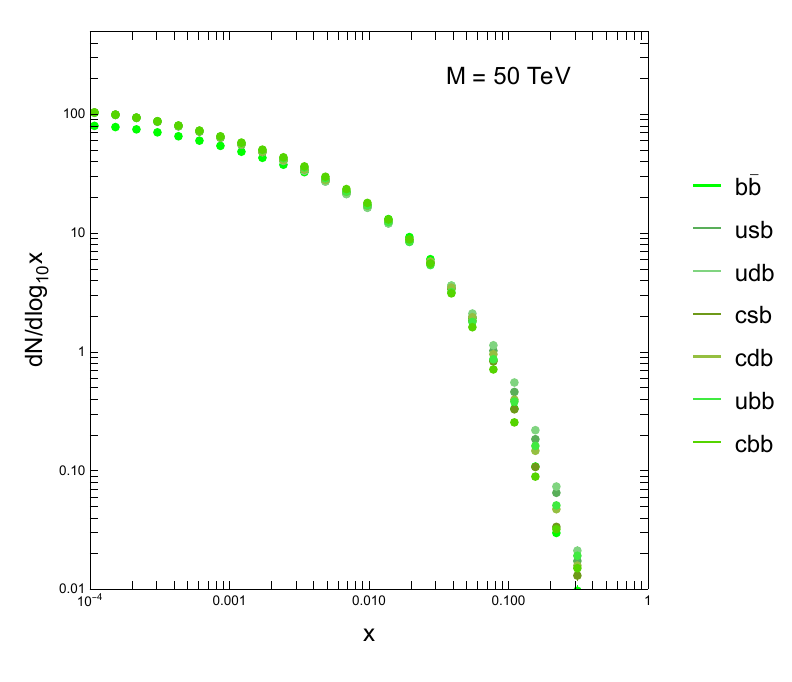}
\end{subfigure}

\vspace{0.5em}

\begin{subfigure}{0.32\textwidth}
    \centering
    \includegraphics[width=\linewidth]{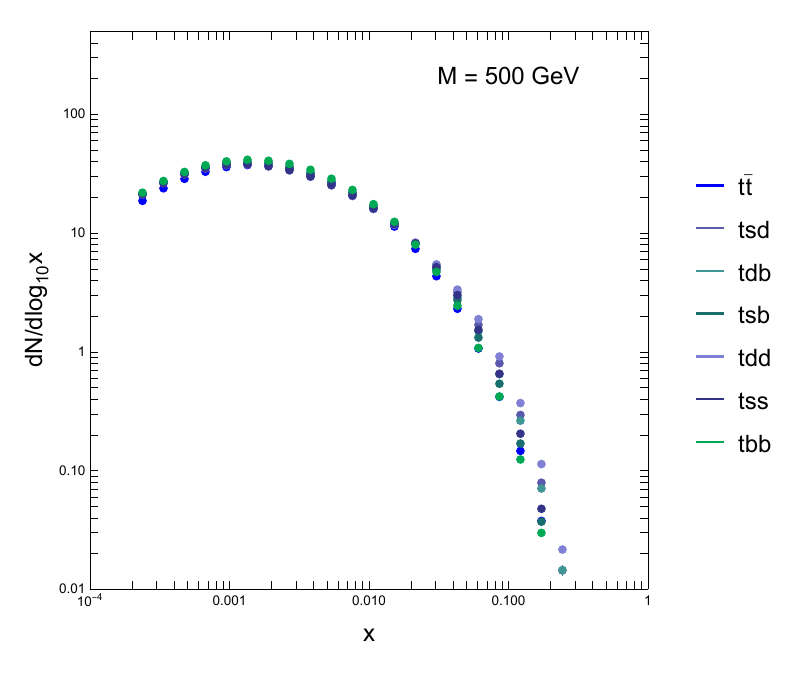}
\end{subfigure}
\hspace*{-1.18cm}
\begin{subfigure}{0.32\textwidth}
    \centering
    \includegraphics[width=\linewidth]{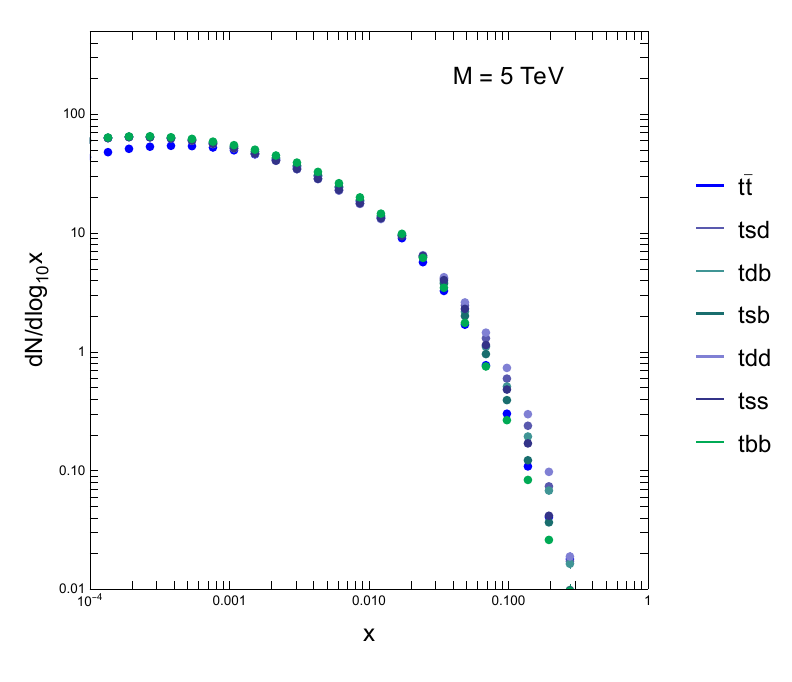}
\end{subfigure}
\hspace*{-1.18cm}
\begin{subfigure}{0.32\textwidth}
    \centering
    \includegraphics[width=\linewidth]{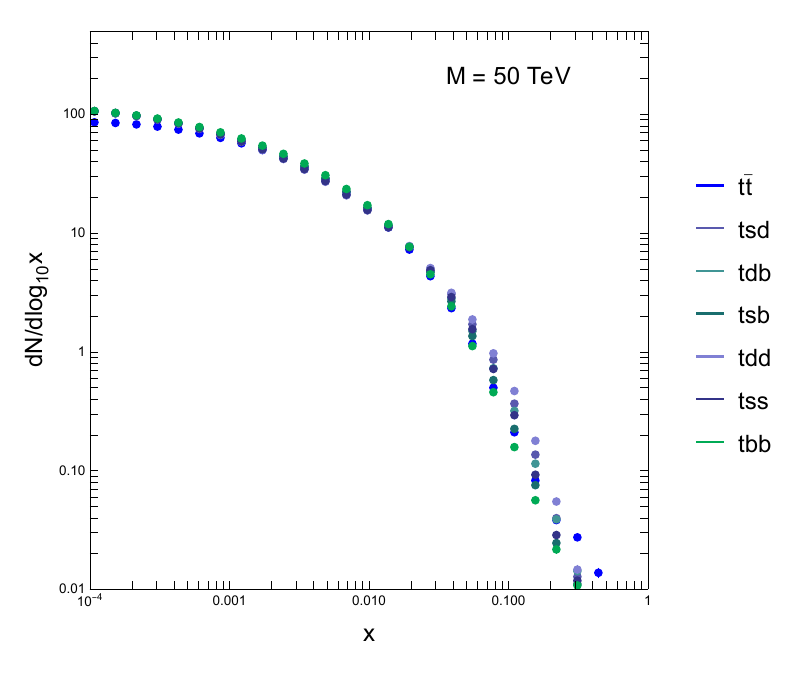}
\end{subfigure}

\caption{Prompt gamma-ray spectra of channels grouped by quark flavor content. From top to bottom, the rows show channels containing only light quarks ($q\bar q$ stands for an equal mixture of $u\bar u$, $d\bar d$ and $s\bar s$); channels containing charm but no bottom or top quarks; channels containing bottom but no top quarks; and channels containing top quarks. The corresponding two-body channels $q\bar q$ (light quarks), $c\bar c$, $b\bar b$, and $t\bar t$ are included in the respective rows for comparison. The three columns correspond to \(M=500\,\mathrm{GeV},\ 5\,\mathrm{TeV},\ 50\,\mathrm{TeV}\), respectively. The spectra show the photon yield per dark-matter decay, $dN_\gamma/d\log_{10}x$ with $x=E_\gamma/M$, obtained by dividing the histogram counts by the number of simulated decays and the logarithmic bin width.}

\label{Fig:gamma_spectra_grid}
\end{figure}

\begin{figure}[t]
\centering

\begin{subfigure}{0.32\textwidth}
    \centering
    \includegraphics[width=\linewidth]{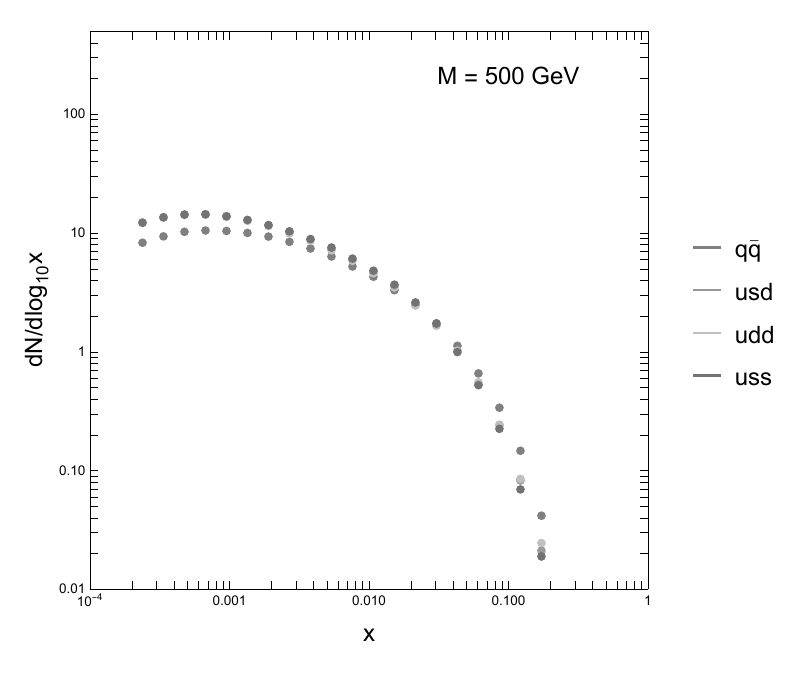}
\end{subfigure}
\hspace*{-1.18cm}
\begin{subfigure}{0.32\textwidth}
    \centering
    \includegraphics[width=\linewidth]{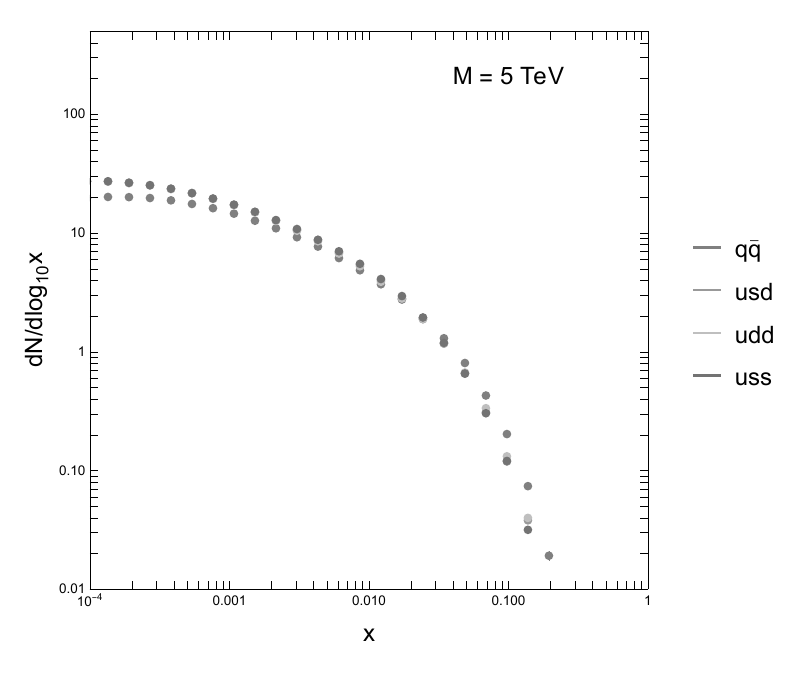}
\end{subfigure}
\hspace*{-1.18cm}
\begin{subfigure}{0.32\textwidth}
    \centering
    \includegraphics[width=\linewidth]{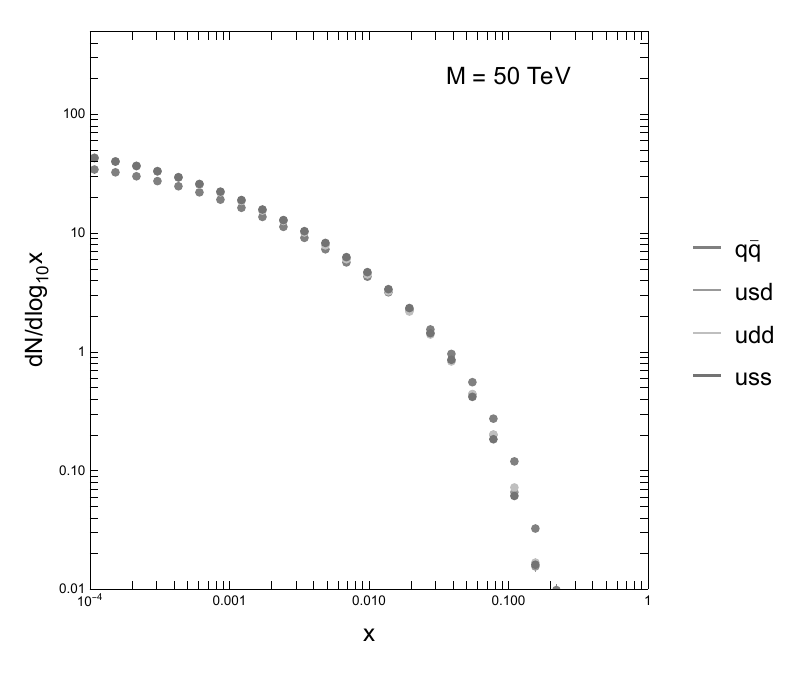}
\end{subfigure}

\vspace{0.5em}

\begin{subfigure}{0.32\textwidth}
    \centering
    \includegraphics[width=\linewidth]{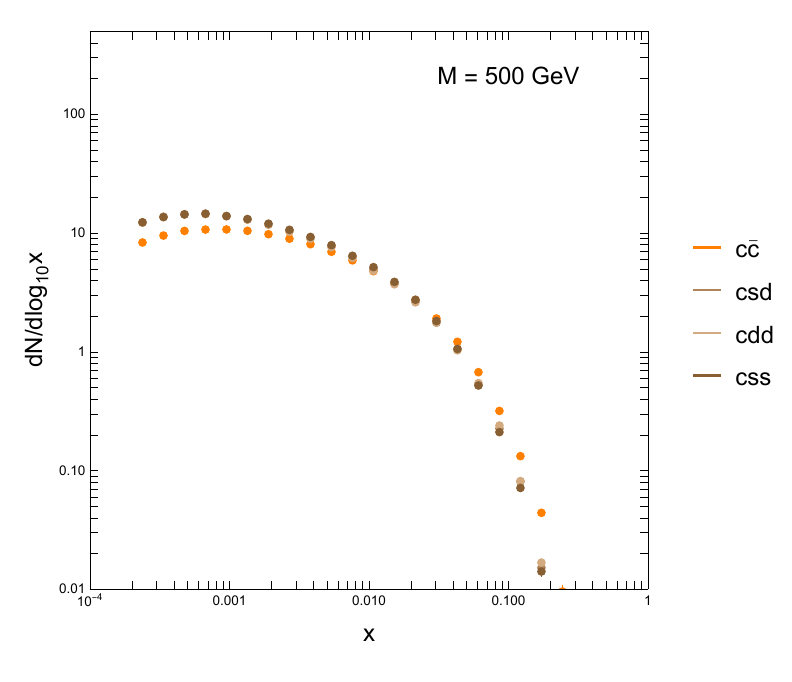}
\end{subfigure}
\hspace*{-1.18cm}
\begin{subfigure}{0.32\textwidth}
    \centering
    \includegraphics[width=\linewidth]{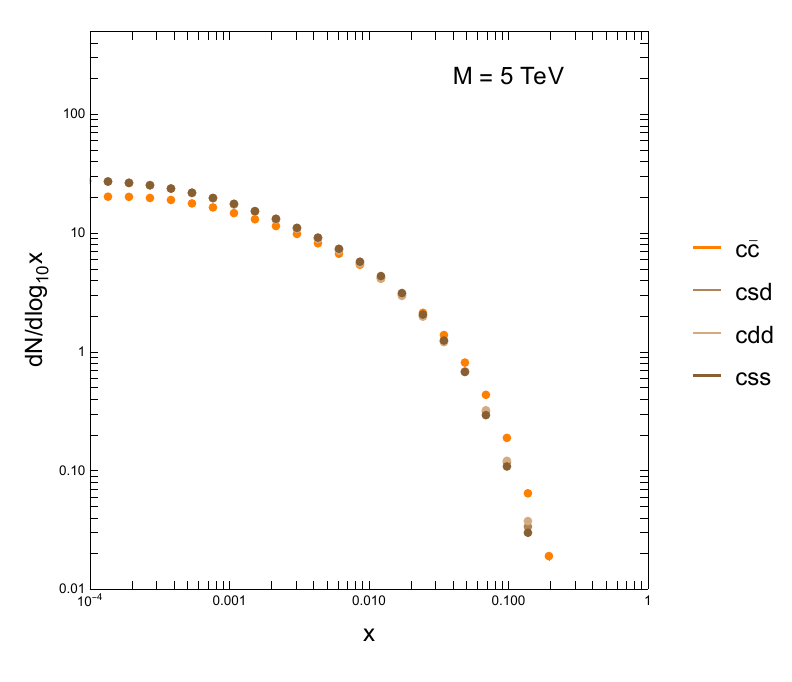}
\end{subfigure}
\hspace*{-1.18cm}
\begin{subfigure}{0.32\textwidth}
    \centering
    \includegraphics[width=\linewidth]{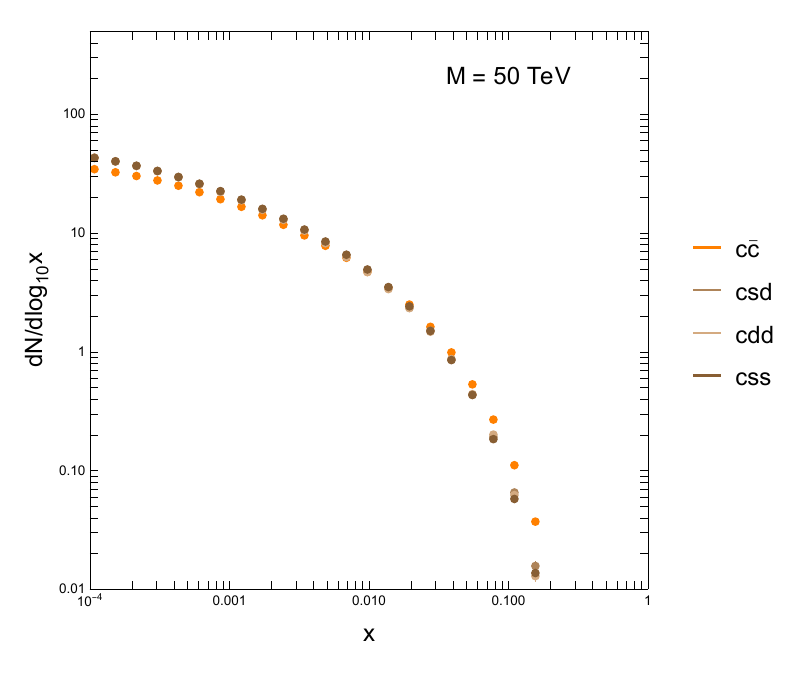}
\end{subfigure}

\vspace{0.5em}

\begin{subfigure}{0.32\textwidth}
    \centering
    \includegraphics[width=\linewidth]{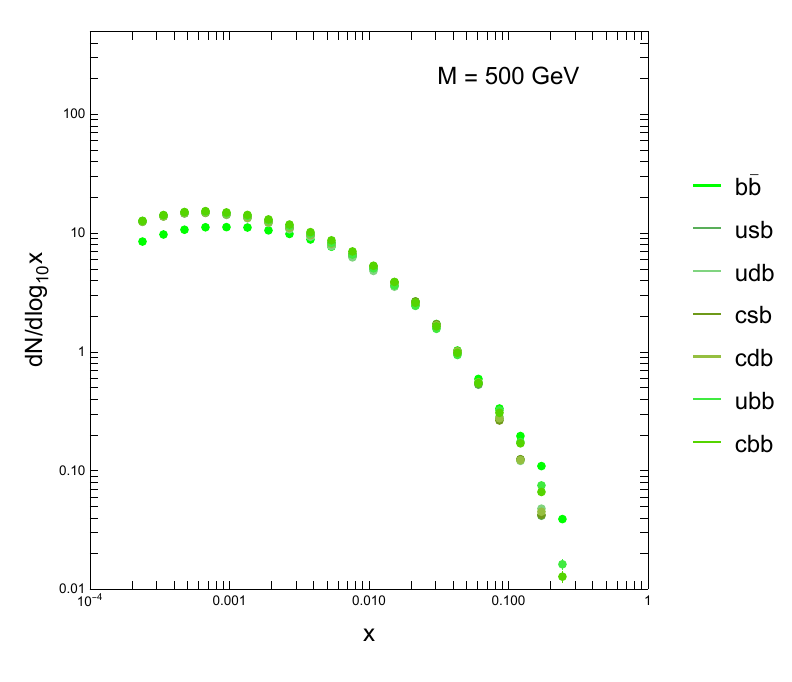}
\end{subfigure}
\hspace*{-1.18cm}
\begin{subfigure}{0.32\textwidth}
    \centering
    \includegraphics[width=\linewidth]{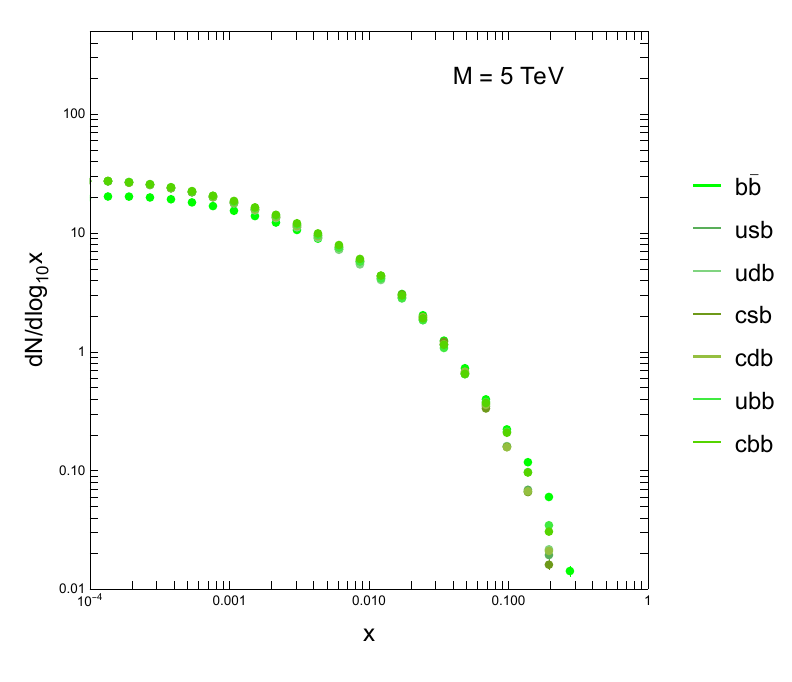}
\end{subfigure}
\hspace*{-1.18cm}
\begin{subfigure}{0.32\textwidth}
    \centering
    \includegraphics[width=\linewidth]{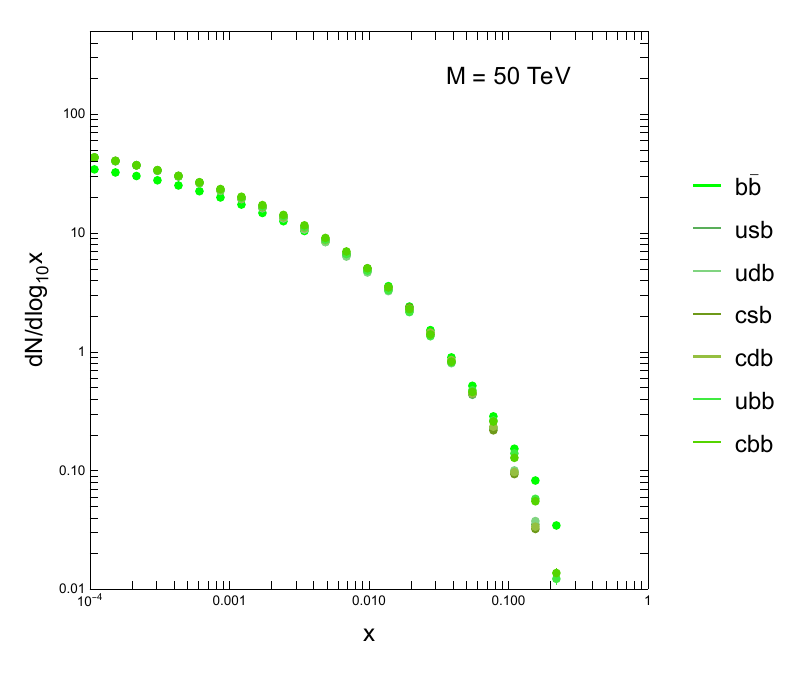}
\end{subfigure}

\vspace{0.5em}

\begin{subfigure}{0.32\textwidth}
    \centering
    \includegraphics[width=\linewidth]{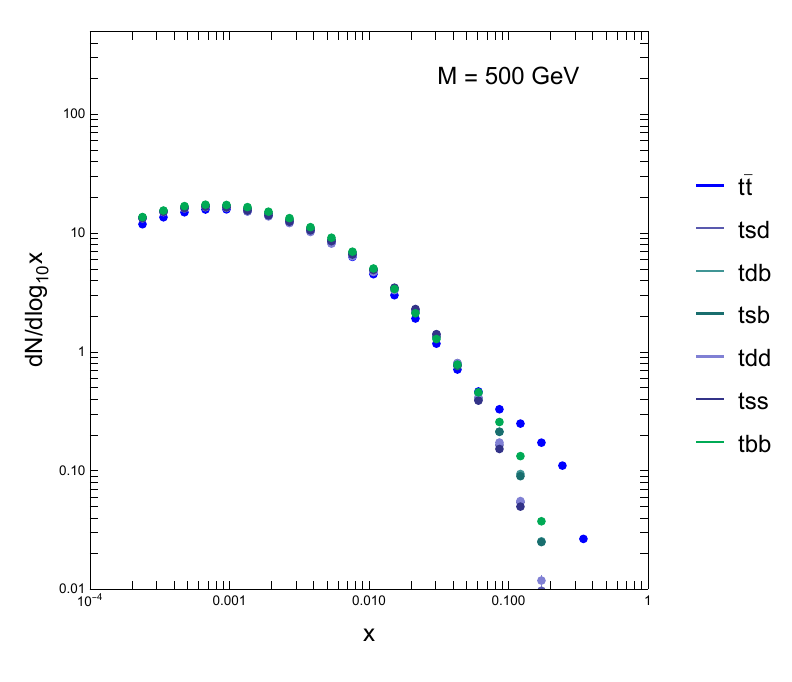}
\end{subfigure}
\hspace*{-1.18cm}
\begin{subfigure}{0.32\textwidth}
    \centering
    \includegraphics[width=\linewidth]{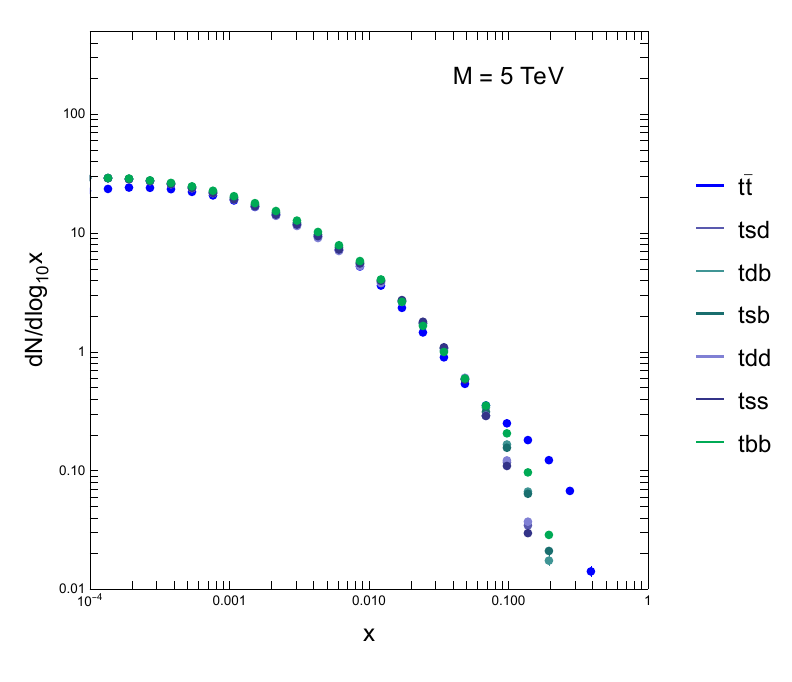}
\end{subfigure}
\hspace*{-1.18cm}
\begin{subfigure}{0.32\textwidth}
    \centering
    \includegraphics[width=\linewidth]{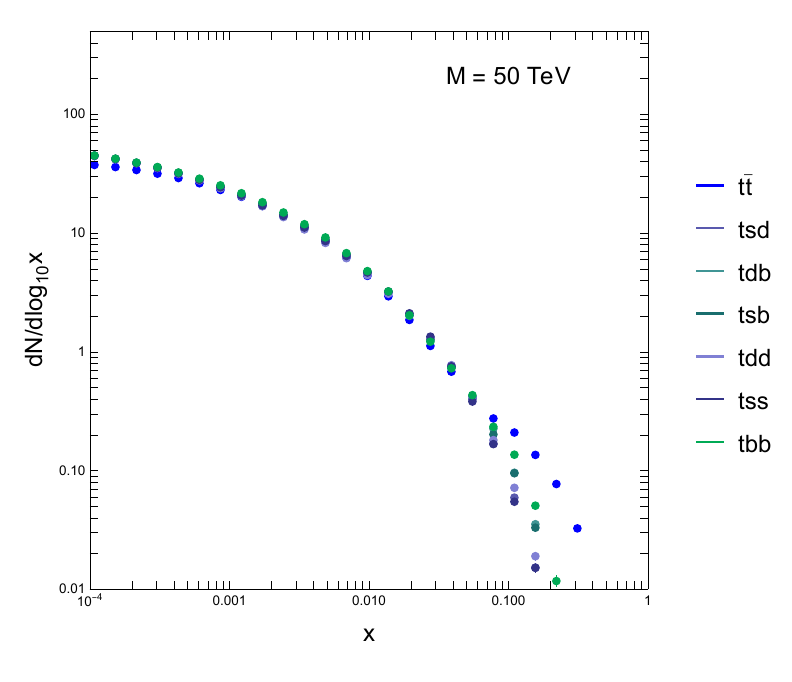}
\end{subfigure}

\caption{Same as figure~\ref{Fig:gamma_spectra_grid}, but for the prompt electron spectra.}
\label{Fig:electron_spectra_grid}
\end{figure}

\begin{figure}[t]
\centering

\begin{subfigure}{0.32\textwidth}
    \centering
    \includegraphics[width=\linewidth]{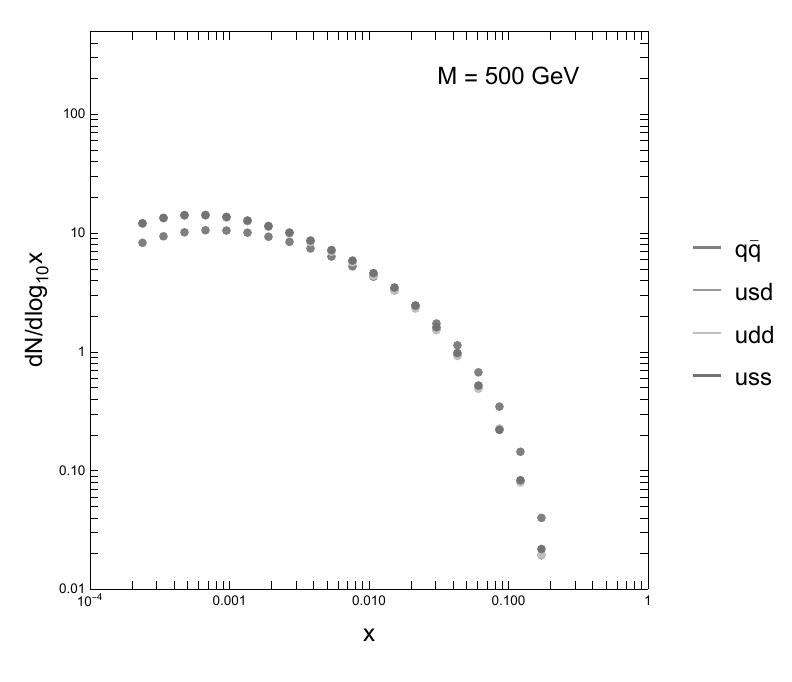}
\end{subfigure}
\hspace*{-1.18cm}
\begin{subfigure}{0.32\textwidth}
    \centering
    \includegraphics[width=\linewidth]{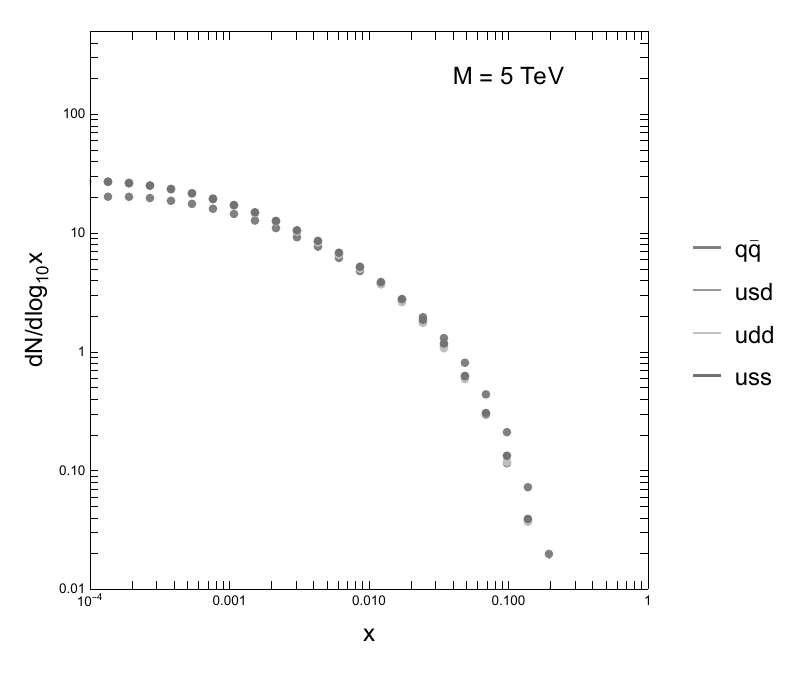}
\end{subfigure}
\hspace*{-1.18cm}
\begin{subfigure}{0.32\textwidth}
    \centering
    \includegraphics[width=\linewidth]{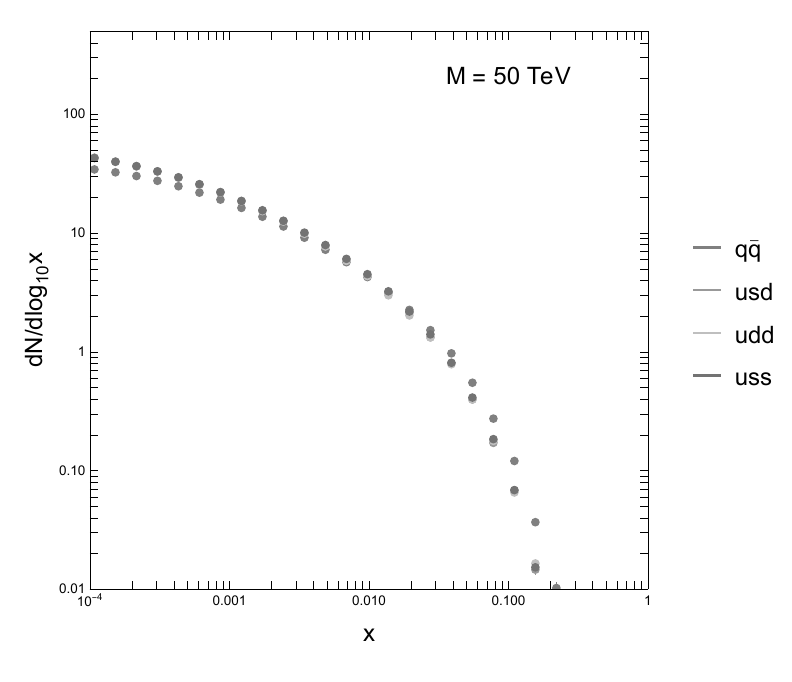}
\end{subfigure}

\vspace{0.5em}

\begin{subfigure}{0.32\textwidth}
    \centering
    \includegraphics[width=\linewidth]{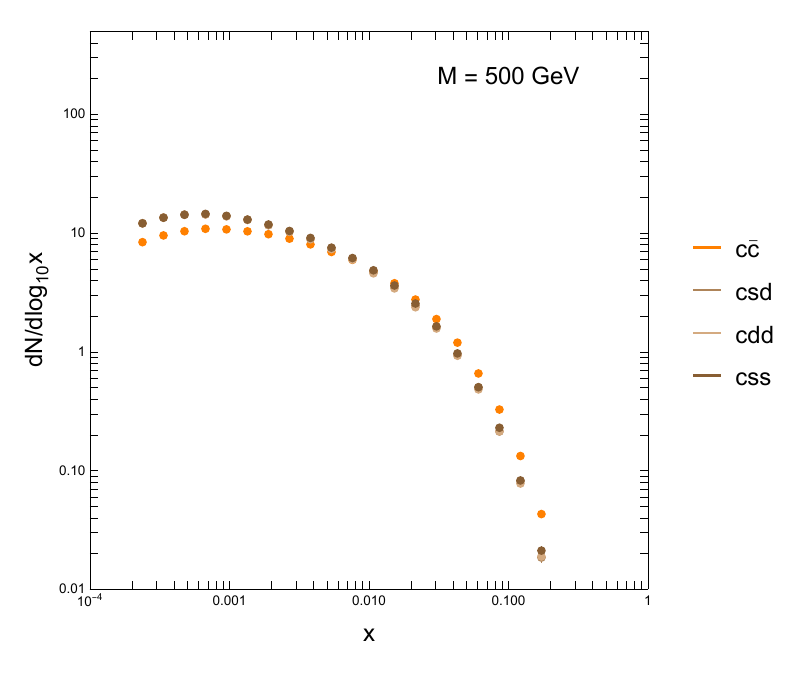}
\end{subfigure}
\hspace*{-1.18cm}
\begin{subfigure}{0.32\textwidth}
    \centering
    \includegraphics[width=\linewidth]{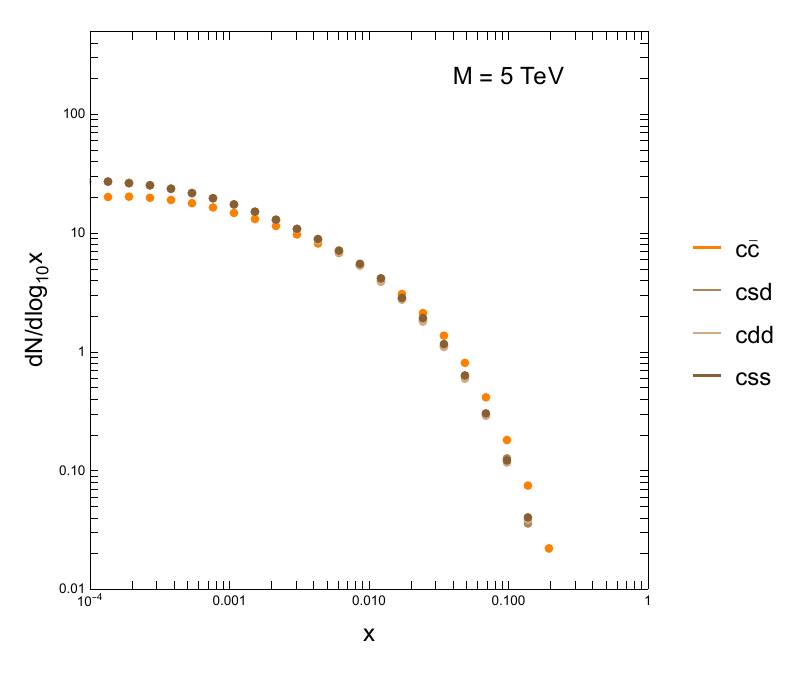}
\end{subfigure}
\hspace*{-1.18cm}
\begin{subfigure}{0.32\textwidth}
    \centering
    \includegraphics[width=\linewidth]{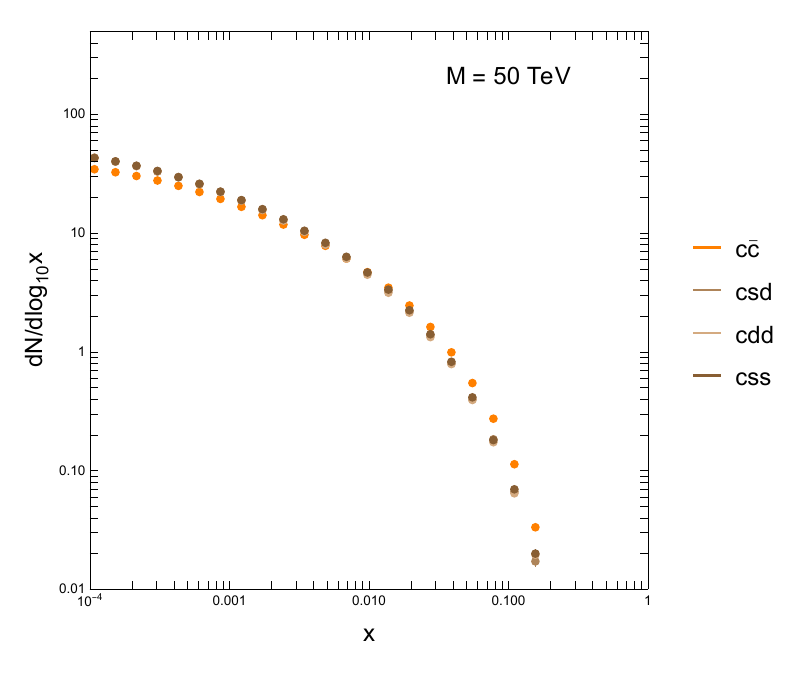}
\end{subfigure}

\vspace{0.5em}

\begin{subfigure}{0.32\textwidth}
    \centering
    \includegraphics[width=\linewidth]{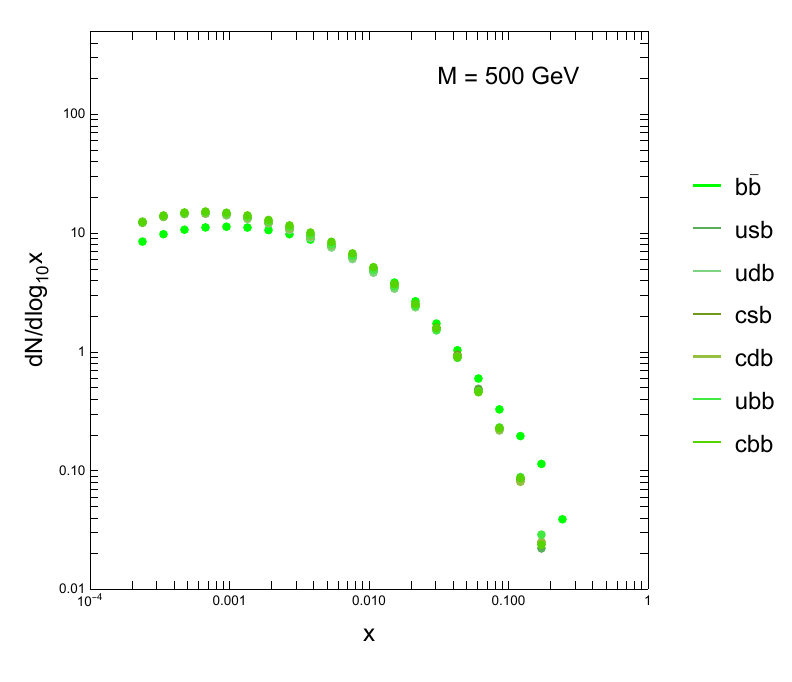}
\end{subfigure}
\hspace*{-1.18cm}
\begin{subfigure}{0.32\textwidth}
    \centering
    \includegraphics[width=\linewidth]{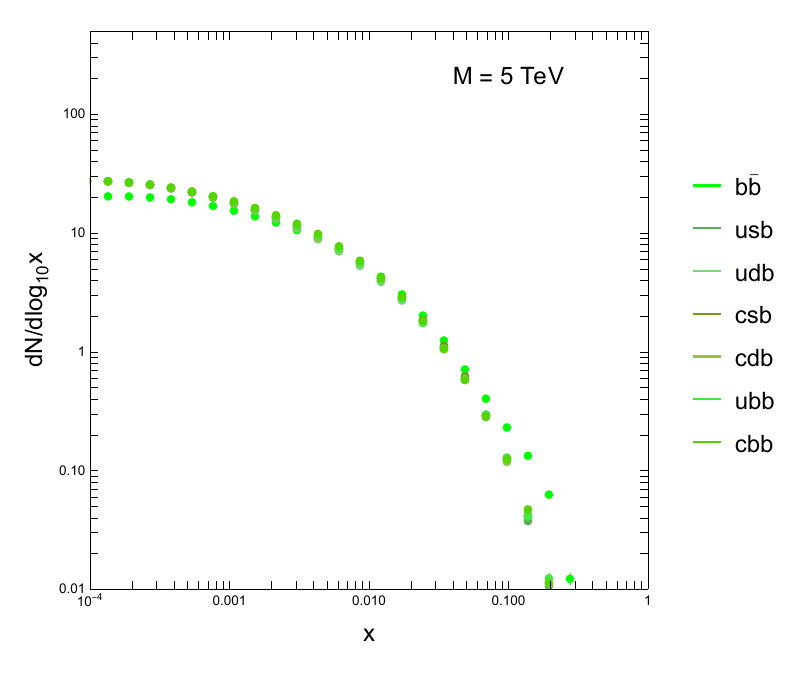}
\end{subfigure}
\hspace*{-1.18cm}
\begin{subfigure}{0.32\textwidth}
    \centering
    \includegraphics[width=\linewidth]{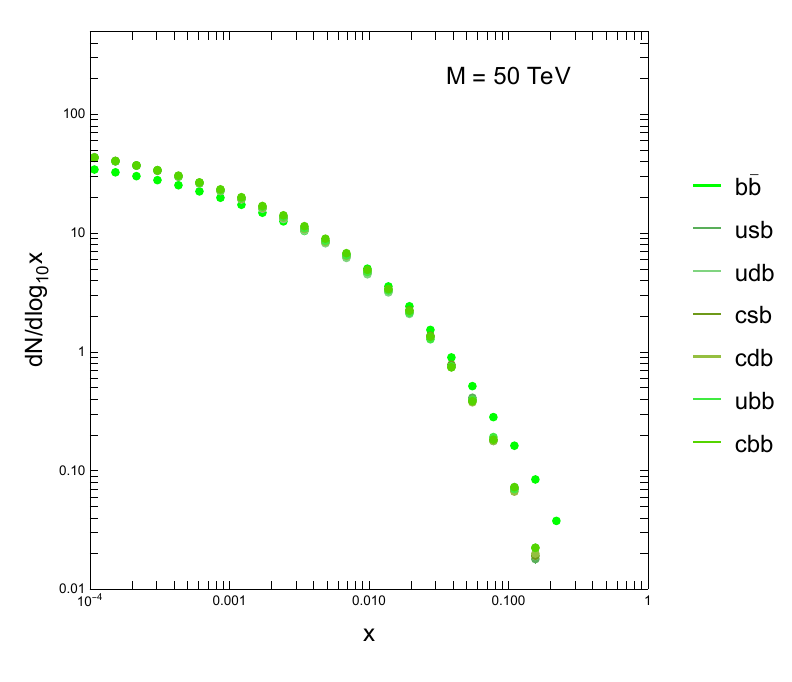}
\end{subfigure}

\vspace{0.5em}

\begin{subfigure}{0.32\textwidth}
    \centering
    \includegraphics[width=\linewidth]{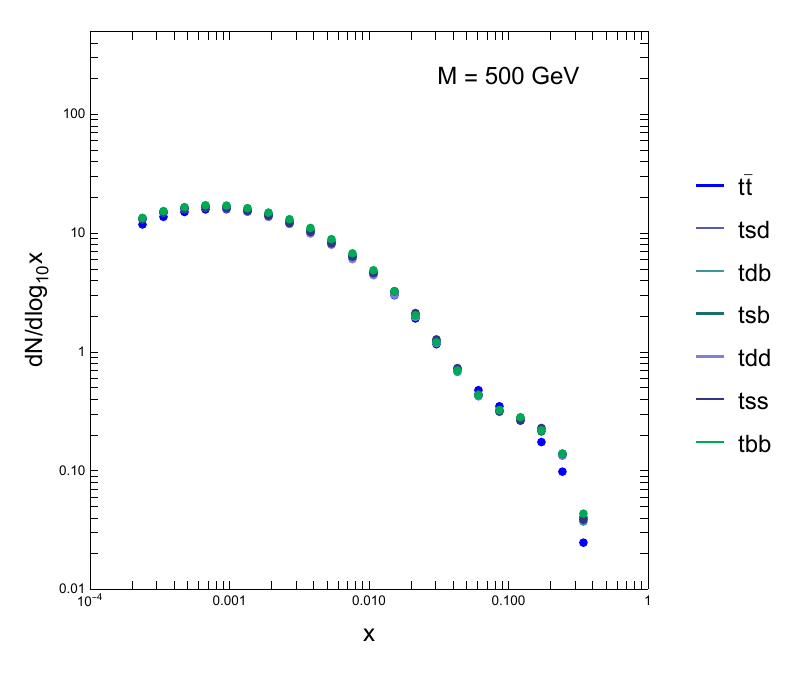}
\end{subfigure}
\hspace*{-1.18cm}
\begin{subfigure}{0.32\textwidth}
    \centering
    \includegraphics[width=\linewidth]{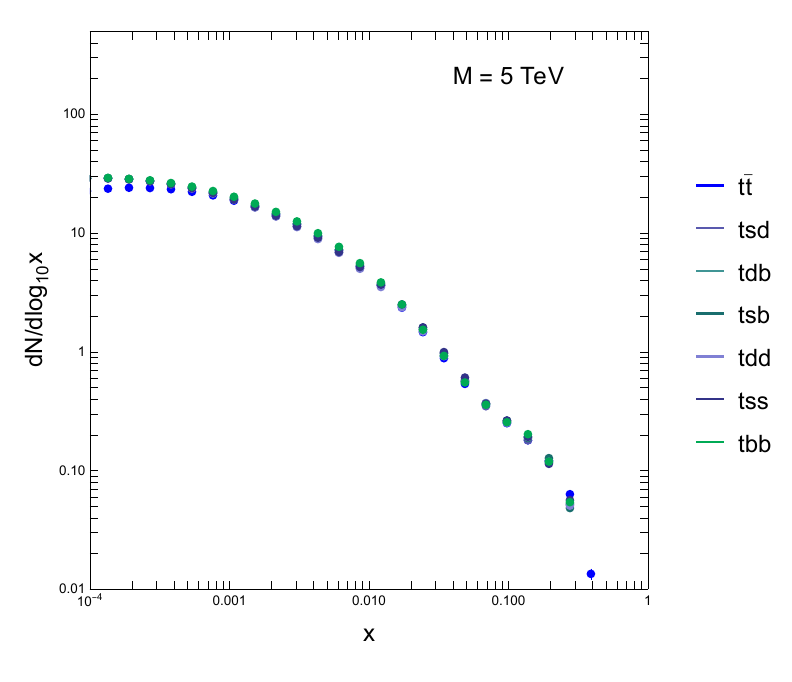}
\end{subfigure}
\hspace*{-1.18cm}
\begin{subfigure}{0.32\textwidth}
    \centering
    \includegraphics[width=\linewidth]{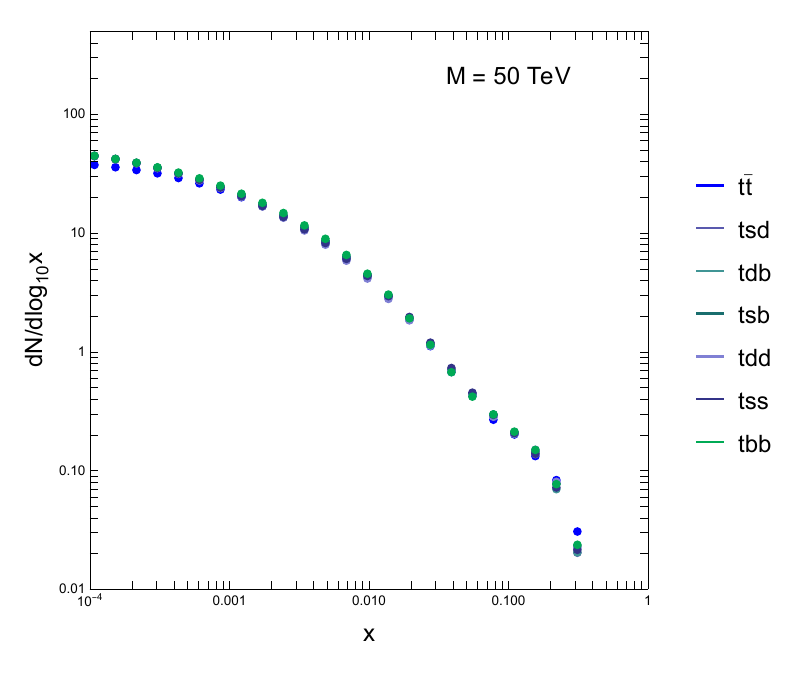}
\end{subfigure}

\caption{Same as figure~\ref{Fig:gamma_spectra_grid}, but for the prompt positron spectra.}
\label{Fig:positron_spectra_grid}
\end{figure}

We choose the DM mass in the range of 10 GeV to 100 TeV, and label the spectra into light quarks $q=(u,d,s)$, charm, bottom, and top categories according to the most massive quark appearing in the final state. The gamma-ray, electron, and positron spectra are shown in  Figures~\ref{Fig:gamma_spectra_grid}, \ref{Fig:electron_spectra_grid}, and \ref{Fig:positron_spectra_grid}. In the figures, we use the energy fraction $x = E/M$ with $M$ being the DM mass, and we show the corresponding spectra as $dN/d\log_{10}x$. We also include the prompt spectrum for the decay of a bosonic DM particle with $B=0$ in each plot for comparison.

The prompt spectra show a significant difference between the $B=0$ and $B=1$ cases. This is mainly due to the two vs. three-body decay kinematics. Three-body decays distribute the available energy among more primary partons, generally shifting the secondary-particle spectra toward lower $x$. In the massless limit, a primary parton can reach $E=M/2$ in either case; the relevant distinction is therefore the spectral weight near the endpoint, rather than the endpoint itself.

\subsection{Effect of Propagation}

The total observable gamma-ray signal from DDM is the sum of four distinct components: prompt emission and secondary inverse Compton scattering (ICS) originating from both the Galactic halo and extragalactic sources.

\subsubsection*{Galactic Contributions}

DDM yield prompt gamma rays through final state radiation, and cascade decays of the final-state quarks, and hadrons. Integrating along a given line of sight in the Galactic halo, the differential flux of prompt gamma rays is given by:
\begin{equation}
\frac{d\Phi_{\text{prompt}}^{\text{Gal}}}{dE_{\gamma}d\Omega}(E_{\gamma},b,l) = \frac{1}{4\pi\tau M} \int_{\text{l.o.s.}} \! ds \, \rho_{\text{NFW}}(r(s; b, l)) \frac{dN_{\gamma}}{dE_{\gamma}}(E_{\gamma})
\end{equation}
where $\tau$ and $M$ are the lifetime and mass of the DM particle, respectively, and $dN_{\gamma}/dE_{\gamma}$ represents the normalized energy spectrum of the final-state photons. We model the DM halo using the standard Navarro-Frenk-White (NFW) density profile, $\rho_{\text{NFW}}(r) = \rho_{s} / [(r/r_{s})(1+r/r_{s})^{2}]$, with a scale density $\rho_{s} = 0.184 \text{ GeV cm}^{-3}$ and a scale radius $r_{s} = 24.42 \text{ kpc}$ as in Refs.~\cite{Navarro:1996gj, Catena:2009mf}. We take the sky-averaged value over the region with $|b| \geq 20^\circ$ and $0 \leq |l| \leq 180^\circ$.\footnote{Some works (Refs.\cite{Cohen:2016uyg,Moulin:2013saa,Liu:2016ngs}) argue that the IGRB represents an isotropic contribution, whereas the Galactic dark matter contribution presents an anisotropic structure, i.e., its intensity depends on the observation direction. As a result, these studies adopt the minimum Galactic contribution, which is located toward the anti-Galactic Center (anti-GC) at $b = 20^\circ$ and a Galactic longitude $l = 180^\circ$. In our analysis, we note the difference between the over-sky average and the anti-GC direction is approximately a factor of 2. The resulting bounds differ by roughly the same factor.}

In addition to prompt photons, final-state electrons and positrons upscatter ambient background photons, such as the cosmic microwave background, starlight, and dust infrared emission, to gamma-ray energies via ICS. The differential flux for this Galactic ICS component is:
\begin{equation}
\frac{d\Phi_{\text{ICS}}^{\text{Gal}}}{dE_{\gamma}d\Omega}(E_{\gamma},b,l) = \frac{1}{E_{\gamma}^{2}}\frac{r_{\odot}}{4\pi\tau} \int_{m_{e}}^{M/2} \!\! dE_{s} \, \frac{\rho_{\odot}}{M} \left( \frac{dN_{e^{+}}}{dE}(E_{s}) + \frac{dN_{e^{-}}}{dE}(E_{s}) \right) I_\mathrm{IC}(E_{\gamma},E_{s},b,l)
\end{equation}
where $r_{\odot} = 8.33 \text{ kpc}$ is the distance to the Galactic Center, and $\rho_{\odot} = 0.3 \text{ GeV cm}^{-3}$ is the local halo density. The numerical propagation function $I_\mathrm{IC}$ accounts for the full Klein-Nishina regime; we adopt a diffusion cylinder half-height of $L = 4 \text{ kpc}$ for this evaluation, consistent with the approach in Ref.~\cite{Blanco:2018esa}. 

\subsubsection*{Extragalactic Contributions}

Analogous decay processes occurring beyond the Milky Way contribute to an isotropic extragalactic flux. The prompt extragalactic differential flux is obtained by integrating over cosmological redshifts:
\begin{equation}
\frac{d\Phi_{\text{prompt}}^{\text{EG}}}{dE_{\gamma}d\Omega}(E_{\gamma}) = \frac{c}{4\pi\tau M} \int_{0}^{1000} \! dz \, \frac{\rho_0}{H(z)} \frac{dN_{\gamma}}{dE_{\gamma}}(E_{\gamma}^{\prime})\, e^{-\tau(E_{\gamma},z)}\,,
\end{equation}
where $\rho_0\simeq1.26\times10^{-6}\,\mathrm{GeV\,cm^{-3}}$, and the Hubble parameter is \[
H(z)=H_0\sqrt{\Omega_m(1+z)^3+(1-\Omega_m)},
\]assuming a flat $\Lambda$CDM cosmology with $\Omega_{m}=0.315$ as in Ref.~\cite{Planck:2018vyg}. We integrate up to $z=1000$, since the contribution from higher redshifts is negligible and this choice improves computational efficiency. The emitted photons are redshifted such that $E_{\gamma}^{\prime} = (1+z)E_{\gamma}$. Furthermore, the factor $e^{-\tau(E_{\gamma},z)}$ accounts for flux attenuation as high-energy gamma rays scatter with the intergalactic ultraviolet (UV) background to produce electron/positron pairs, which we model under the minimal UV background assumption, following the treatments discussed in Refs.~\cite{Cirelli:2010xx, Dominguez:2010bv}. 

Finally, intergalactic electrons and positrons upscatter CMB photons, generating the extragalactic ICS component. We neglect their propagation over cosmological distances and assume that they radiate at the redshift at which they are injected. The resulting flux is
\begin{equation}
\begin{aligned}
\frac{d\Phi_{\rm ICS}^{\rm EG}}{dE_\gamma d\Omega}(E_\gamma)
&=\frac{c\rho_0}{4\pi\tau M E_\gamma}\,
j_{\rm IC}^{\rm CMB}(E_\gamma,0)
\int_0^{1000}\!dz\,
\frac{e^{-\tau(E_\gamma,z)}}{(1+z)^2H(z)},
\end{aligned}
\end{equation}
where $\rho_0\simeq1.26\times10^{-6}\,\mathrm{GeV\,cm^{-3}}$ and $j_{\rm IC}^{\rm CMB}(E_\gamma,0)$ is the effective inverse Compton emission kernel at $z=0$, defined by 
\begin{equation}
\begin{aligned}
j_{\rm IC}^{\rm CMB}(E_\gamma,0)
&\equiv \int_{m_e}^{M/2}\!dE_e\,
\frac{P_{\rm IC}^{\rm CMB}(E_\gamma,E_e,0)}
     {b_{\rm IC}^{\rm CMB}(E_e,0)}\,\mathcal I_e(E_e),    
\end{aligned}
\end{equation}
Here, $\mathcal I_e(E_e)$ is the integrated electron and positron yield above $E_e$:

\begin{equation}
\begin{aligned}
\mathcal I_e(E_e)
&\equiv \int_{E_e}^{M/2}\!dE_s\,
\left[\frac{dN_{e^+}}{dE_s}(E_s)+\frac{dN_{e^-}}{dE_s}(E_s)\right].
\end{aligned}
\end{equation}
In the Thomson approximation, the differential inverse Compton power
at $z=0$ is
\begin{equation}
P_{\rm IC}^{\rm CMB}(E_\gamma,E_e,0)
=
\frac{3\sigma_T}{4\gamma^2}E_\gamma
\int_0^1 \! \frac{dy}{y}\,
n_{\rm CMB}\!\left(\frac{E_\gamma}{4\gamma^2y},0\right)
\left[2y\ln y+y+1-2y^2\right],
\end{equation}
while the electron energy-loss rate is
\begin{equation}
b_{\rm IC}^{\rm CMB}(E_e,0)
=
\frac{4\sigma_T}{3m_e^2}E_e^2u_{\gamma,\rm CMB}(0),
\end{equation}
following Refs.~\cite{Cirelli:2009vg, Meade:2009iu}.
Here, $\gamma=E_e/m_e$, $\sigma_T$ is the Thomson cross section, $n_{\rm CMB}(\epsilon,z)$ is the CMB photon number density per unit energy, and $u_{\gamma,\rm CMB}(z)$ is the CMB energy density. We take $T_{\rm CMB}(z)=2.7255(1+z)\,\mathrm{K}$. 

For this differential-power convention, the blackbody scaling gives
\begin{equation}
\begin{aligned}
P_{\rm IC}^{\rm CMB}((1+z)E_\gamma,E_e,z)
&=(1+z)^3P_{\rm IC}^{\rm CMB}(E_\gamma,E_e,0),\\
b_{\rm IC}^{\rm CMB}(E_e,z)
&=(1+z)^4b_{\rm IC}^{\rm CMB}(E_e,0).
\end{aligned}
\end{equation}
These scalings give a factor of $(1+z)^{-1}$ in the ratio
$P_{\rm IC}^{\rm CMB}/b_{\rm IC}^{\rm CMB}$. Converting the differential power to a photon-number spectrum requires division by the emitted photon energy, $E_\gamma'=(1+z)E_\gamma$, providing the additional factor
of $(1+z)^{-1}$ and hence the $(1+z)^{-2}$ factor in the extragalactic ICS flux above.

\begin{figure}[h!]
    \centering

    \begin{subfigure}{0.38\textwidth}
        \centering
        \includegraphics[width=\linewidth]{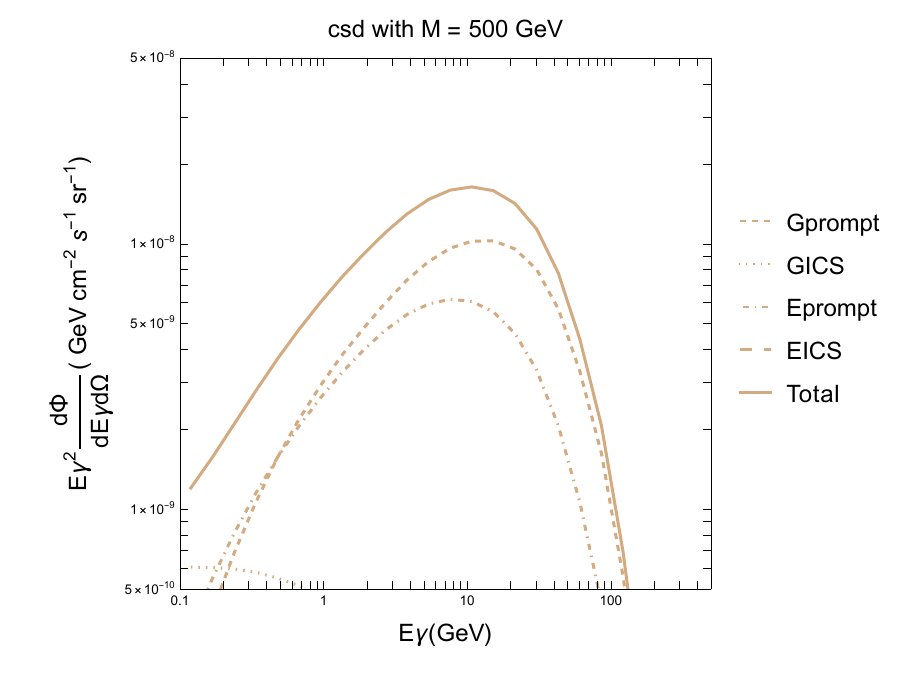}
    \end{subfigure}
    \hspace*{-1.45cm}
    \begin{subfigure}{0.38\textwidth}
        \centering
        \includegraphics[width=\linewidth]{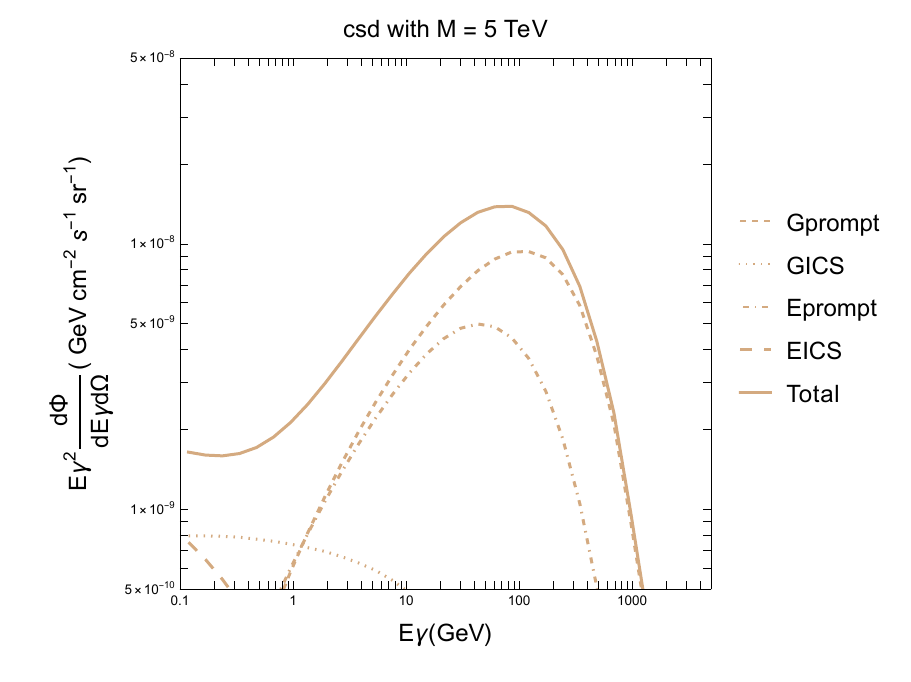}
    \end{subfigure}
    \hspace*{-1.45cm}
    \begin{subfigure}{0.38\textwidth}
        \centering
        \includegraphics[width=\linewidth]{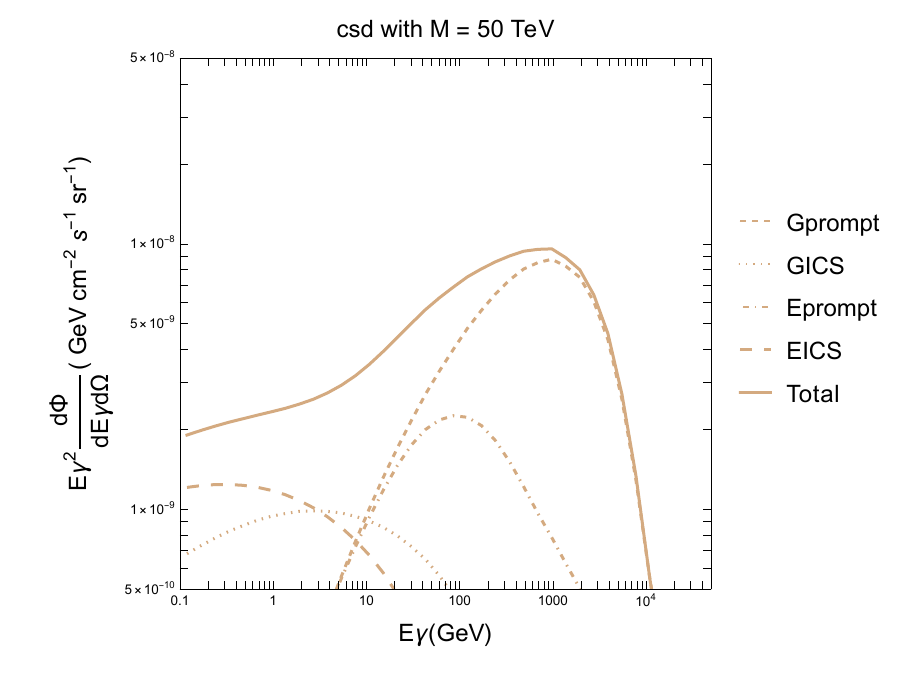}
    \end{subfigure}

    \caption{The components of the total gamma-ray flux from $\chi\to csd$ for three different masses:~\(M=500\,\mathrm{GeV},\ 5\,\mathrm{TeV},\ 50\, \mathrm{TeV}\), and \(\tau=10^{28}\,\mathrm{s}\) is used for normalization. The solid line represents the total flux, the dashed line represents the galactic prompt component, the dotted line represents galactic ICS, the short dash-dotted line represents the extragalactic prompt component, and the long dash-dotted line represents extragalactic ICS.}
    \label{Fig: XCSD component comparison}
\end{figure}

\begin{figure}[h!]
    \centering
    \includegraphics[scale=0.8]{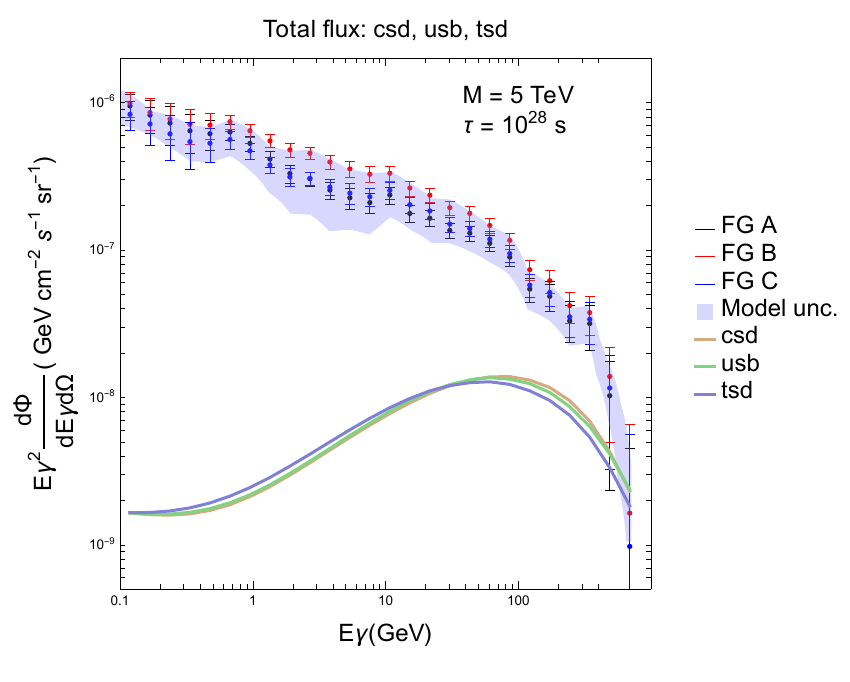}
\caption{Differential flux of gamma rays for DDM ($M = 5\,\mathrm{TeV}$, $\tau = 1 \times 10^{28}\,\mathrm{s}$) and the IGRB for three foreground models. The signal is shown for three representative flavor channels: $csd$, $bsu$, and $tsd$. The light-blue band indicates the IGRB modeling uncertainty. See the main text for details.}
\label{Fig: IGRB and flux}
\end{figure}

The sum of these four components yields the total expected gamma-ray flux, as illustrated in Figure~\ref{Fig: XCSD component comparison}.\footnote{Extragalactic electromagnetic cascades arise from repeated pair production and ICS \cite{Blanco:2018esa, Blanco:2018GammaCascade}. We omit this contribution from our baseline calculation. Tests with the public \texttt{gamma-Cascade} code for representative channels ($\chi\to csd$ over $0.5$--$100\,\mathrm{TeV}$, and $\chi\to udd$, $tbb$, and $q\bar q$ at $50$ and $100\,\mathrm{TeV}$) indicate only small changes in the lifetime bounds, although the cascade contribution can be substantial in the lowest-energy bins at high masses. These estimates depend on the assumed extragalactic background light and intergalactic magnetic fields.} The high-energy endpoints of the total spectra are dominated by prompt gamma rays, while secondary gamma rays contribute significantly at lower energies. Figure~\ref{Fig: IGRB and flux} compares the signal spectra for a few flavor combinations to the IGRB for a particular benchmark value of $M$ and $\tau$. 

\section{Statistical Procedure for Limit Setting}
\label{Isotropic Gamma-Ray Background and Statistical Methods}

Having established the expected gamma-ray fluxes in our model, we are in a position to use the IGRB, as measured by the Fermi-LAT collaboration~\cite{Fermi-LAT:2014ryh}, and constrain the DM lifetime. The IGRB represents the residual diffuse emission after the removal of known, resolved sources and the subtraction of modeled Galactic foregrounds. In this section, we detail the statistical methodologies used to translate these observations into $3\sigma$ exclusion limits on the DM lifetime. To provide a comprehensive evaluation, we derive these bounds under three distinct frameworks: a conservative approach that assumes no astrophysical background, a phenomenological background model with a combined power-law and exponential spectrum, and a shape-based likelihood analysis that explicitly models individual astrophysical contributors to the IGRB, such as star-forming galaxies and AGN. All three analyses retain the full set of 26 IGRB energy bins.

\subsection{IGRB foreground modeling}

To isolate the IGRB, the Fermi-LAT collaboration removes known sources, subtracts the modeled diffuse emissions, masks the bright GC region, and only includes observations with Galactic latitudes above 20 degrees. The precise determination of the diffuse emission foreground is model-dependent. Therefore, the data is categorized using three distinct foreground models, labeled A, B, and C, which are derived based on different assumptions about cosmic rays. Specifically, Model A assumes that cosmic ray sources trace the Galactic pulsar distribution with constant diffusion and re-acceleration parameters throughout the Galaxy. Model B introduces an additional electron-only source population near the Galactic center that is responsible for the majority of the inverse Compton emission. Finally, Model C adopts a cosmic ray source distribution more centrally peaked than pulsars, alongside diffusion and re-acceleration parameters that vary with Galactocentric radius and height. Further details regarding these models can be found in Ref.~\cite{Fermi-LAT:2014ryh}.

Figure~\ref{Fig: IGRB and flux} shows the differential IGRB flux for these three foregrounds, as well as the signal for decaying dark matter for three representative final states with $M = 5\,\mathrm{TeV}$ and $\tau = 1 \times 10^{28}\,\mathrm{s}$. For each energy bin, $\sigma_i$ is taken from the published Fermi-LAT IGRB spectral error bar. These errors include the statistical uncertainty, the LAT effective-area parametrization uncertainty, and the uncertainty from residual cosmic-ray background subtraction, added in quadrature. The light blue band encapsulates the foreground modeling uncertainty, which is not folded into $\sigma_i$. 

\subsection{Statistical Procedure}

To establish robust constraints on the dark matter lifetime, we evaluate the bounds using three complementary frameworks: a background-free approach, a phenomenological background model, and a detailed astrophysical likelihood analysis. 

\subsubsection*{Background-Free (BGF) Constraints}

The first and most conservative approach assumes that no astrophysical sources contribute to the observed IGRB. We take $\phi_i$ and $\sigma_i$ to be the intensity and corresponding uncertainty in bin $i$ for foreground model B.\footnote{This definition differs from those in Refs.~\cite{Fermi-LAT:2015qzw, Liu:2016ngs}, where the value of $\phi_i$ is obtained by incrementing the foreground A flux by its $1\sigma$ modeling uncertainty. We have checked that the resulting difference in the BGF bounds is within $10\%$.} With this assumption, only bins where the DM flux exceeds the adopted IGRB upper intensity contribute to the test statistic:
\begin{equation}
\chi^2_{\mathrm{BGF}} \;=\;
\sum_{\,i \in \{\, i \mid \phi_i^{\mathrm{DM}} > \phi_i \,\}}
\frac{\bigl[\phi_i^{\mathrm{DM}} - \phi_i \bigr]^2}{\sigma_i^{2}}\,.
\end{equation}
This is basically a one-sided upper-limit criterion, used in both Refs.~\cite{Fermi-LAT:2015qzw,Liu:2016ngs}. In each bin, $\sigma_{i}$ includes the statistical and instrumental systematic uncertainties represented by the error bars in Figure~\ref{Fig: IGRB and flux}, but not the Galactic foreground modeling uncertainty (illustrated by the light blue band). 
We also leave out the uncertainties associated with the MC based signal prediction as these are much smaller in comparison.

\begin{figure}[h!]
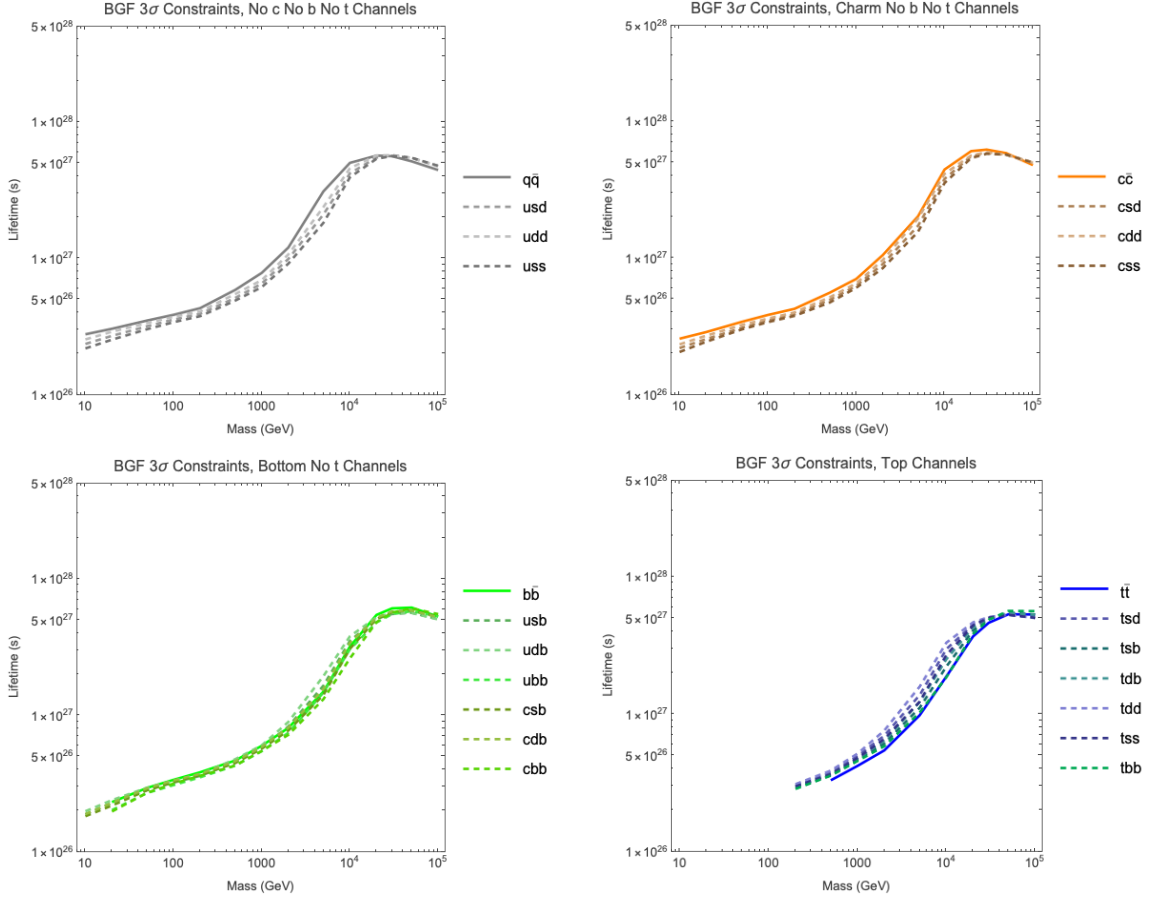

\centering

\begin{subfigure}{0.48\textwidth}
    \centering
    \includegraphics[width=\linewidth]{figures/constraints_conservative_no_c_no_b_no_t.pdf}
\end{subfigure}
\hfill
\begin{subfigure}{0.48\textwidth}
    \centering
    \includegraphics[width=\linewidth]{figures/constraints_conservative_c_no_b_no_t.pdf}
\end{subfigure}

\vspace{0.5em}

\begin{subfigure}{0.48\textwidth}
    \centering
    \includegraphics[width=\linewidth]{figures/constraints_conservative_bottom_no_t.pdf}
\end{subfigure}
\hfill
\begin{subfigure}{0.48\textwidth}
    \centering
    \includegraphics[width=\linewidth]{figures/constraints_conservative_top.pdf}
\end{subfigure}

\caption{\(3\sigma\) BGF limits grouped by the quark flavor content of the final state: only light quarks (\(u,d,s\), upper left); charm but no bottom or top quarks (upper right); bottom but no top quarks (lower left); and top quarks (lower right).}
\label{Fig: background-free bounds}
\end{figure}

As a function of $M$, we scan the lifetime $\tau$ and determine the 3$\sigma$ limit based on the condition $\chi^2_{\rm BGF}=9$. Our results are shown in Figure~\ref{Fig: background-free bounds}. 

\subsubsection*{Power-Law Exponential (PLE) Model}

For our second approach we adopt a phenomenological background model with a power-law plus exponential (PLE) energy dependence, following Refs.~\cite{Fermi-LAT:2015qzw, Liu:2016ngs}. This corresponds to an assumption that unresolved astrophysical sources are responsible for the overall spectral shape of the IGRB. Constraints on the dark matter signal are derived by injecting the expected dark matter contribution on top of a phenomenological best-fit background model. We define
\begin{equation}
\chi^2_{\mathrm{PLE}} \;=\;
\sum_i
\frac{\bigl[\phi_i - \phi_i^{\mathrm{bkg}}(I,\gamma,E_{\rm cut}) - \phi_i^{\mathrm{DM}}\bigr]^2}{\sigma_i^{2}}\,,
\label{equation: sensitive limits}
\end{equation}
where the astrophysical background flux, $\phi_{i}^{\mathrm{bkg}}$, is modeled as a power-law with an exponential cutoff\footnote{In Ref.~\cite{Fermi-LAT:2014ryh}, power law and broken power law models are also considered, but the power-law plus exponential cutoff provides the best fit.}
\begin{equation}
\frac{dN}{dE}
= I\left(\frac{E}{100\,\mathrm{MeV}}\right)^{-\gamma}
  \exp\!\left(-\frac{E}{E_{\mathrm{cut}}}\right).
\end{equation}
We adopt the best-fit parameters ($I$, $\gamma$, and $E_{cut}$) derived for each foreground model by the Fermi-LAT collaboration~\cite{Fermi-LAT:2014ryh}, as summarized in Table~\ref{tab:ple_params}. As a contrast to the very conservative approach of the BGF method, for the PLE method we base the limits on $\tau$ on foreground model C, which generally yields the strongest bounds.

\begin{table}[h!]
\centering
\renewcommand{\arraystretch}{1.2}
\begin{tabular}{lccc}
\hline\hline
Foreground Model & $I$ ($10^{-7} \, \text{MeV}^{-1}\text{cm}^{-2}\text{s}^{-1}\text{sr}^{-1}$) & $\gamma$ & $E_{\text{cut}}$ (GeV) \\
\hline
Model A & $0.95 \pm 0.08$ & $2.32 \pm 0.02$ & $279 \pm 52$ \\
Model B & $1.12 \pm 0.08$ & $2.28 \pm 0.02$ & $206 \pm 31$ \\
Model C & $0.78 \pm 0.07$ & $2.26 \pm 0.02$ & $233 \pm 41$ \\
\hline\hline
\end{tabular}
\caption{Best-fit parameters for the IGRB power-law plus exponential cutoff (PLE) background model, adopted from Ref.~\cite{Fermi-LAT:2014ryh}.}
\label{tab:ple_params}
\end{table}

\begin{figure}[h!]
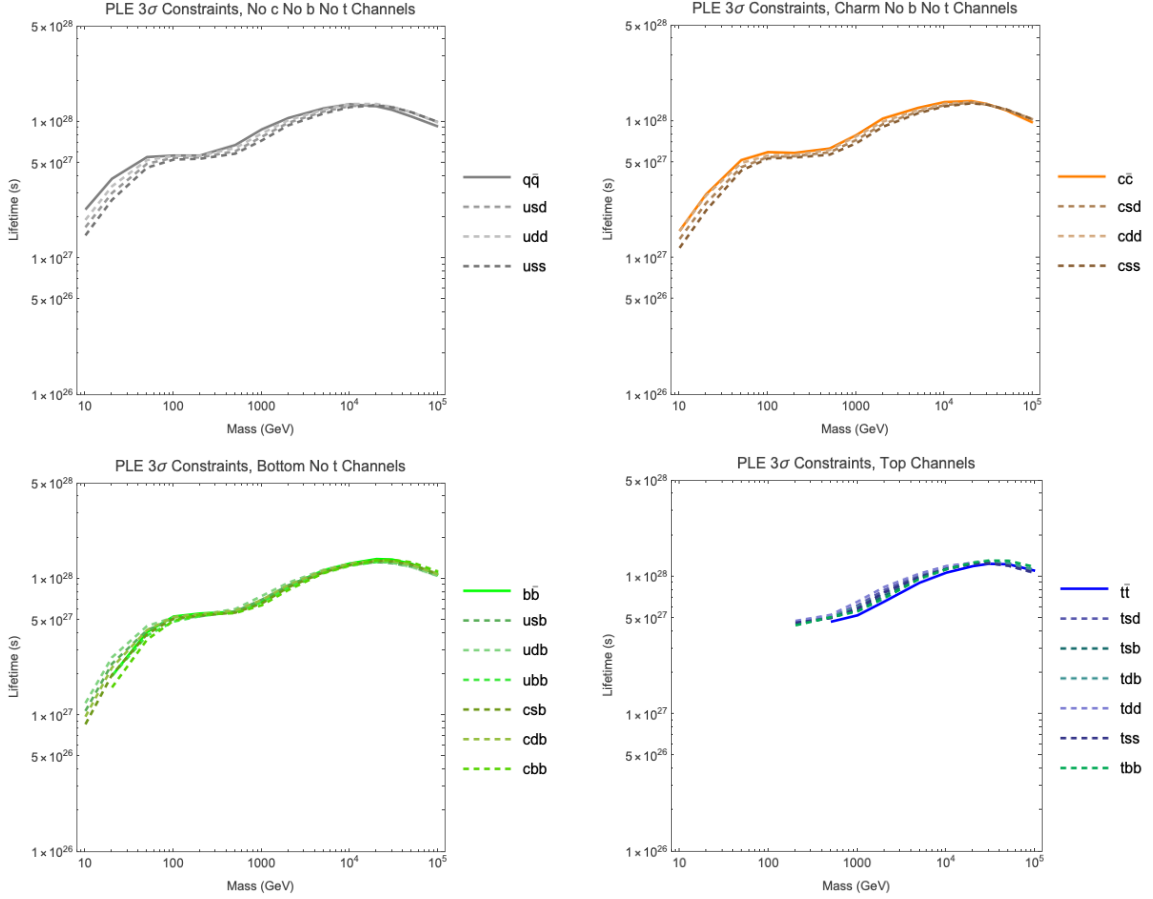

\centering

\begin{subfigure}{0.48\textwidth}
    \centering
    \includegraphics[width=\linewidth]{figures/constraints_sensitive_C_no_c_no_b_no_t.pdf}
\end{subfigure}
\hfill
\begin{subfigure}{0.48\textwidth}
    \centering
    \includegraphics[width=\linewidth]{figures/constraints_sensitive_C_c_no_b_no_t.pdf}
\end{subfigure}

\vspace{0.5em}

\begin{subfigure}{0.48\textwidth}
    \centering
    \includegraphics[width=\linewidth]{figures/constraints_sensitive_C_bottom_no_t.pdf}
\end{subfigure}
\hfill
\begin{subfigure}{0.48\textwidth}
    \centering
    \includegraphics[width=\linewidth]{figures/constraints_sensitive_C_top.pdf}
\end{subfigure}

\caption{\(3\sigma\) PLE limits with foreground C, grouped by the quark flavor content of the final state: only light quarks (\(u,d,s\), upper left); charm but no bottom or top quarks (upper right); bottom but no top quarks (lower left); and top quarks (lower right).}
\label{Fig: PLE bounds}
\end{figure}

To determine the DM lifetime limits, we fix the PLE parameters to the foreground-C values in Table~\ref{tab:ple_params} throughout the lifetime scan. Using all 26 IGRB energy bins, the background-only model gives $\chi_0^2=11.2188$. At each DM mass $M$, we determine the $3\sigma$ lifetime limit by finding the value of $\tau$ for which $\Delta\chi^2=9$. Our results are shown in Figure~\ref{Fig: PLE bounds}.  \footnote{We show PLE limits for foreground model C, which generally yields stronger bounds than models A and B. For the likelihood analysis, we show the strongest limit among foreground models A, B, and C at each channel and mass.}

\subsubsection*{Likelihood Method}

\begin{figure}[h!]
    \centering
    \includegraphics
    {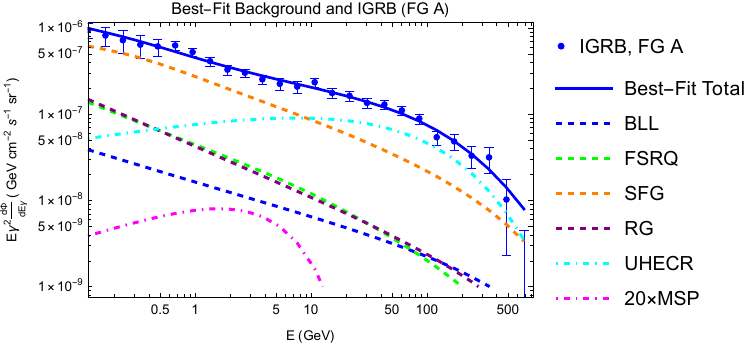}
\caption{Individual background components contributing to the IGRB in the likelihood-based method, including BLL, FSRQ, SFG, RG, UHECR, and MSP (scaled by 20 in the plot). The blue solid line presents the best-fit total background. The blue bins and error bars correspond to the IGRB derived   from foreground model A.}
\label{Figure: background_analysis}
\end{figure}

While the PLE background model is simple, it is phenomenological and likely leads to a limit that is too aggressive, as it does not allow for degeneracies between the presence of signal and subcomponents of the gamma ray background in the observed shape of the observed flux. In order to derive limits that are more nuanced, potentially weaker than the PLE approach but more reliable, we introduce our third and most sophisticated method to constrain the DM lifetime. In doing so, we incorporate details from recent studies for modeling the astrophysical components of the gamma ray background~\cite{Blanco:2018esa, Fornasa:2015qua, Cholis:2024hmd, Blanco:2021icw}. In particular, Ref.~\cite{Cholis:2024hmd} incorporates these distinct populations into a comprehensive likelihood analysis. Extensive multi-wavelength observations have demonstrated that the IGRB is dominated by a combination of unresolved active galactic nuclei, such as BL Lac objects (BLLs), flat-spectrum radio quasars (FSRQs), and radio galaxies (RGs), alongside star-forming galaxies (SFGs)~\cite{Fornasa:2015qua, Korsmeier:2022cwp, Ajello:2020zna, Gruppioni:2013jna, Blanco:2021icw, Linden:2016fdd}. Truly diffuse processes, such as cascades from ultra-high-energy cosmic rays (UHECRs), and emission from millisecond pulsars (MSPs), also contribute at a modest level~\cite{Koers:2008ba, Siegal-Gaskins:2010oks, Cholis:2024hmd}.

The total gamma-ray intensity from these populations is evaluated by integrating their empirical luminosity functions and energy spectra over cosmological redshifts, accounting for flux attenuation due to pair production on the extragalactic background light. Rather than detailing the standard multi-wavelength correlations and parameterizations here, we adopt the publicly available flux templates from Ref.~\cite{Cholis:2024hmd}. We also follow the normalization and spectral-shape nuisance ranges given in Ref.~\cite{Cholis:2024hmd}, minimizing $\chi^2$ over these nuisance parameters for the background-only (null) hypothesis, with the results shown in Figure~\ref{Figure: background_analysis}, yielding a best-fit value of $\chi^{2} = 12.7$\footnote{This corresponds to $\chi^2/\mathrm{dof}=0.79$, as reported in Ref.~\cite{Cholis:2024hmd}} for foreground A, and $\chi^2=17.0$ and $10.9$ for foregrounds B and C, respectively.

\begin{figure}[h!]
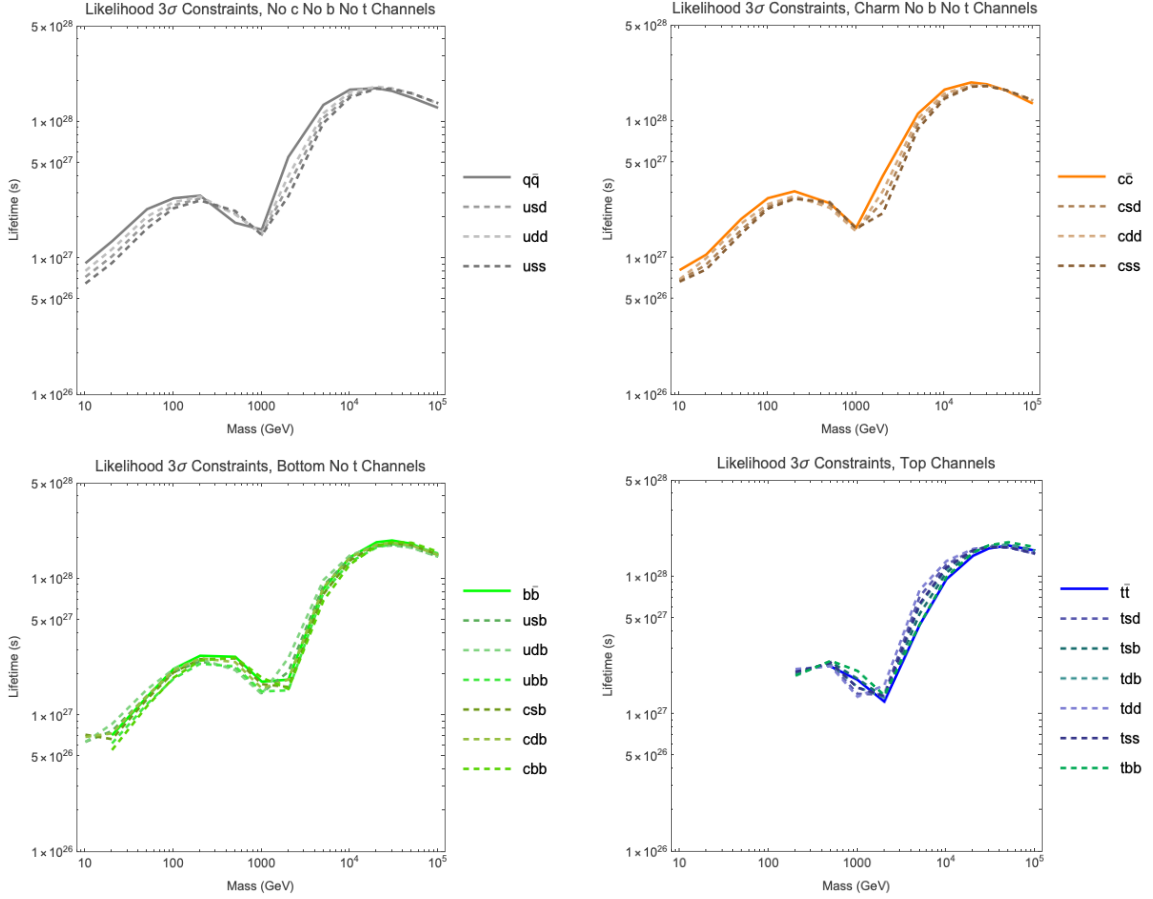

\centering

\begin{subfigure}{0.48\textwidth}
    \centering
    \includegraphics[width=\linewidth]{figures/constraints_likelihood_A_no_c_no_b_no_t.pdf}
\end{subfigure}
\hfill
\begin{subfigure}{0.48\textwidth}
    \centering
    \includegraphics[width=\linewidth]{figures/constraints_likelihood_A_c_no_b_no_t.pdf}
\end{subfigure}

\vspace{0.5em}

\begin{subfigure}{0.48\textwidth}
    \centering
    \includegraphics[width=\linewidth]{figures/constraints_likelihood_A_bottom_no_t.pdf}
\end{subfigure}
\hfill
\begin{subfigure}{0.48\textwidth}
    \centering
    \includegraphics[width=\linewidth]{figures/constraints_likelihood_A_top.pdf}
\end{subfigure}

\caption{\(3\sigma\) likelihood limits taking the strongest bound among foregrounds A, B, and C at each channel and mass, grouped by the quark flavor content of the final state: only light quarks (\(u,d,s\), upper left); charm but no bottom or top quarks (upper right); bottom but no top quarks (lower left); and top quarks (lower right).}
\label{Fig: likelihood bounds}
\end{figure}

For each $(M,\tau)$, we add the DM signal and minimize $\chi^2$ over the background nuisance parameters, including the UHECR normalization. We determine the $3\sigma$ lifetime limit from $\Delta\chi^2=9$, measured relative to the signal-free best fit. The resulting bounds are shown in Figure~\ref{Fig: likelihood bounds}. 

\subsection{Discussion of Results}
\label{Results}
The $3\sigma$ exclusion limits on the dark matter lifetime for all decay channels under the BGF, PLE, and likelihood frameworks are shown in Figures~\ref{Fig: background-free bounds}, \ref{Fig: PLE bounds}, and~\ref{Fig: likelihood bounds}, respectively.\footnote{We omit the $M=10\,\mathrm{GeV}$ point for $\chi\to b\bar b$, since it lies below the open-bottom threshold $2m_B\simeq10.56\,\mathrm{GeV}$. The $b\bar b$ bounds therefore start at $M=20\,\mathrm{GeV}$.}

Across all three statistical frameworks, there are some common patterns. The ranges of lifetime bounds for three-body decays of $B=1$ DDM and two-body decays of conventional $B=0$ DDM are generically similar. Still, within each statistical framework, the ratio of a $B=1$ bound to the corresponding $B=0$ bound shown in the figures ranges from about $0.52$ to $1.80$, depending on the DM mass and flavor channel. Weaker $B=1$ bounds in some channels may be explained by the larger spectral weight near the high-energy end in two-body decays, and at the highest DM masses (above  $30\,\mathrm{TeV}$) the corresponding bounds agree within about $10\%$. An example of a stronger $B=1$ bound is the comparison of a final state like $tbb$ in the $B=1$ case vs. $t\bar{t}$ in the $B=0$ case for lower DM masses, due to a simple kinematic reason: near threshold, two top quarks reduce the available kinetic energy in the event, and therefore the $B=0$ spectrum shifts down to lower energies compared to the $B=1$ case, which only has a single top quark in the final state. We would however expect that limits for a flavor violating decay channel such as $\chi\to tc$ in the $B=0$ case would still result in stronger bounds compared to $B=1$ decay channels containing a top quark in the final state.

Comparing the different final states of $\mathrm{B}=1$ DDM to each other, over most of the mass range below tens of TeV, the light-flavor channel $\chi\to udd$ exhibits the strongest bounds, whereas the heavy-flavor channel $\chi\to tbb$ is generally among the more weakly constrained channels in this mass range. This trend is generally followed by the other flavor combinations: channels with lighter quarks tend to give stronger bounds in this mass range, while those involving heavier quarks lead to weaker bounds. This trend also exists in the decay of $\mathrm{B}=0$ DM: the lighter-quark channels exhibit the stronger bounds, whereas the heavier-quark channels yield the weaker constraints below about $10\,\mathrm{TeV}$. 

In contrast, for masses larger than tens of TeV, the trend generally reverses, with the lighter quark final states tending to yield weaker bounds. The trend can be interpreted as follows: at lower DM masses, the bound is driven by the spectrum near the endpoint, which is higher for the lighter flavor final states. As $M$ increases, the Fermi-LAT energy range probes progressively lower $x=E/M$, and the bounds become more sensitive to the lower-$x$ photons and electrons produced through showering and subsequent decays. One can observe a similar trend for $\mathrm{B}=0$ DDM.

Finally, comparing the three statistical methodologies, we observe that the BGF limits provide the weakest bounds on the DM lifetime as expected. The PLE limits lead to the strongest bounds below $5\,\mathrm{TeV}$, at the cost of relying on the phenomenological background spectrum without a deeper justification. The likelihood method results in somewhat weaker bounds compared to the PLE method at these masses, but is arguably more reliable due to the more sophisticated background modeling, which allows there to be mild degeneracies between different combinations of background subcomponents and the signal in fitting the observed shape of the IGRB. At $10\,\mathrm{TeV}$ the likelihood limits are stronger than PLE for most channels, and from $20\,\mathrm{TeV}$ upward for all channels, by about $20$--$40\%$.

\section{Conclusion}
\label{Conclusion}

In this work, we explored the indirect detection signatures of DDM with $B=1$, specifically in gamma rays. If the DM particle is a SM gauge singlet fermion, it is expected to decay into three quarks via the ``baryon portal''. We calculated the expected prompt and secondary gamma-ray spectra for all eighteen allowed flavor combinations kinematically accessible over a DM mass range of $10\,\mathrm{GeV}$ to $100\,\mathrm{TeV}$. By comparing these predictions against Fermi-LAT measurements of the IGRB, we derived $3\sigma$ exclusion limits on the DM lifetime using three statistical frameworks: a conservative background-free approach, a phenomenologically motivated power-law plus exponential background model, and a likelihood-based shape analysis.

Our results indicate that light-quark final states generally yield stronger lifetime bounds at lower masses, while heavy-quark final states can yield comparable or stronger bounds at masses of tens of TeV and above. This change reflects the increasing importance of the lower-$x$ part of the signal spectrum within the Fermi-LAT energy range. Comparing our statistical methods, the detailed likelihood analysis provides a physical middle ground between the overly conservative BGF and the arguably too aggressive PLE methods below $5\,\mathrm{TeV}$; while it yields the strongest bounds for the majority of final states at $10\,\mathrm{TeV}$ and all final states from $20\,\mathrm{TeV}$ upward. 

Finally, we have demonstrated that the exclusion limits for the $B=1$ DDM model, while being generically similar to the limits for the $B=0$ case, can differ from them by about a factor of 2, depending on the mass and the final state, in each of the three statistical frameworks. Only at the highest masses of $M\geq30\,\mathrm{TeV}$, the limits become equal for all practical purposes.

These results highlight the importance of using channel-specific three-body spectra when constraining $B=1$ DDM, since applying two-body bounds directly can either overestimate or underestimate the excluded parameter space.
\acknowledgments{

The authors thank Kimberly Boddy for helpful discussions and feedback.
The research of CK and ZY was funded by the National Science Foundation Grant Number PHY-2210562 during the initial stages of this work. TY is supported by the Gravity, Spacetime, and Particle Physics (GRASP) Initiative from Harvard University, the NSF grant PHY-2309456 and the Samsung Science and Technology Foundation under Project Number SSTF-BA2201-06.
}

\appendix

\bibliographystyle{JHEP}
\bibliography{bibtex}{}

@article{Blanco:2018GammaCascade,
    author = "Blanco, Carlos",
    title = "{$\gamma$-Cascade: A Simple Program to Compute Cosmological Gamma-Ray Propagation}",
    eprint = "1804.00005",
    archivePrefix = "arXiv",
    primaryClass = "astro-ph.HE",
    doi = "10.1088/1475-7516/2019/01/013",
    journal = "JCAP",
    volume = "01",
    pages = "013",
    year = "2019"
}

@article{Alloul:2013bka,
    author = "Alloul, Adam and Christensen, Neil D. and Degrande, C{\'e}line and Duhr, Claude and Fuks, Benjamin",
    title = "{FeynRules  2.0 - A complete toolbox for tree-level phenomenology}",
    eprint = "1310.1921",
    archivePrefix = "arXiv",
    primaryClass = "hep-ph",
    reportNumber = "CERN-PH-TH-2013-239, MCNET-13-14, IPPP-13-71, DCPT-13-142, PITT-PACC-1308",
    doi = "10.1016/j.cpc.2014.04.012",
    journal = "Comput. Phys. Commun.",
    volume = "185",
    pages = "2250--2300",
    year = "2014"
}

@article{Degrande:2011ua,
    author = "Degrande, Celine and Duhr, Claude and Fuks, Benjamin and Grellscheid, David and Mattelaer, Olivier and Reiter, Thomas",
    title = "{UFO - The Universal FeynRules Output}",
    eprint = "1108.2040",
    archivePrefix = "arXiv",
    primaryClass = "hep-ph",
    reportNumber = "CP3-11-25, IPHC-PHENO-11-04, IPPP-11-39, DCPT-11-78, MPP-2011-68",
    doi = "10.1016/j.cpc.2012.01.022",
    journal = "Comput. Phys. Commun.",
    volume = "183",
    pages = "1201--1214",
    year = "2012"
}

@article{Alwall:2014hca,
    author = "Alwall, J. and Frederix, R. and Frixione, S. and Hirschi, V. and Maltoni, F. and Mattelaer, O. and Shao, H. -S. and Stelzer, T. and Torrielli, P. and Zaro, M.",
    title = "{The automated computation of tree-level and next-to-leading order differential cross sections, and their matching to parton shower simulations}",
    eprint = "1405.0301",
    archivePrefix = "arXiv",
    primaryClass = "hep-ph",
    reportNumber = "CERN-PH-TH-2014-064, CP3-14-18, LPN14-066, MCNET-14-09, ZU-TH-14-14",
    doi = "10.1007/JHEP07(2014)079",
    journal = "JHEP",
    volume = "07",
    pages = "079",
    year = "2014"
}

@article{Navarro:1996gj,
    author = "Navarro, Julio F. and Frenk, Carlos S. and White, Simon D. M.",
    title = "{A Universal density profile from hierarchical clustering}",
    eprint = "astro-ph/9611107",
    archivePrefix = "arXiv",
    doi = "10.1086/304888",
    journal = "Astrophys. J.",
    volume = "490",
    pages = "493--508",
    year = "1997"
}

@article{Catena:2009mf,
    author = "Catena, Riccardo and Ullio, Piero",
    title = "{A novel determination of the local dark matter density}",
    eprint = "0907.0018",
    archivePrefix = "arXiv",
    primaryClass = "astro-ph.CO",
    doi = "10.1088/1475-7516/2010/08/004",
    journal = "JCAP",
    volume = "08",
    pages = "004",
    year = "2010"
}

@article{Planck:2018vyg,
    author = "Aghanim, N. and others",
    collaboration = "Planck",
    title = "{Planck 2018 results. VI. Cosmological parameters}",
    eprint = "1807.06209",
    archivePrefix = "arXiv",
    primaryClass = "astro-ph.CO",
    doi = "10.1051/0004-6361/201833910",
    journal = "Astron. Astrophys.",
    volume = "641",
    pages = "A6",
    year = "2020",
    note = "[Erratum: Astron.Astrophys. 652, C4 (2021)]"
}

@article{Fermi-LAT:2015qzw,
    author = "Ackermann, M. and others",
    collaboration = "Fermi-LAT",
    title = "{Limits on Dark Matter Annihilation Signals from the Fermi LAT 4-year Measurement of the Isotropic Gamma-Ray Background}",
    eprint = "1501.05464",
    archivePrefix = "arXiv",
    primaryClass = "astro-ph.CO",
    reportNumber = "FERMILAB-PUB-15-052-A",
    doi = "10.1088/1475-7516/2015/09/008",
    journal = "JCAP",
    volume = "09",
    pages = "008",
    year = "2015"
}

@article{Liu:2016ngs,
    author = "Liu, Wei and Bi, Xiao-Jun and Lin, Su-Jie and Yin, Peng-Fei",
    title = "{Constraints on dark matter annihilation and decay from the isotropic gamma-ray background}",
    eprint = "1602.01012",
    archivePrefix = "arXiv",
    primaryClass = "astro-ph.CO",
    doi = "10.1088/1674-1137/41/4/045104",
    journal = "Chin. Phys. C",
    volume = "41",
    number = "4",
    pages = "045104",
    year = "2017"
}

@article{Cholis:2024hmd,
    author = "Cholis, Ilias and Krommydas, Iason",
    title = "{Scrutinizing the isotropic gamma-ray background in search of dark matter}",
    eprint = "2408.11421",
    archivePrefix = "arXiv",
    primaryClass = "astro-ph.HE",
    doi = "10.1103/PhysRevD.110.103032",
    journal = "Phys. Rev. D",
    volume = "110",
    number = "10",
    pages = "103032",
    year = "2024"
}

@article{Blanco:2018esa,
    author = "Blanco, Carlos and Hooper, Dan",
    title = "{Constraints on Decaying Dark Matter from the Isotropic Gamma-Ray Background}",
    eprint = "1811.05988",
    archivePrefix = "arXiv",
    primaryClass = "astro-ph.HE",
    reportNumber = "FERMILAB-PUB-18-627-A",
    doi = "10.1088/1475-7516/2019/03/019",
    journal = "JCAP",
    volume = "03",
    pages = "019",
    year = "2019"
}

@inproceedings{Moulin:2013saa,
    author = "Moulin, E. and Cirelli, M. and Panci, P. and Serpico, P. and Viana, A.",
    title = "{Gamma-ray constraints on decaying Dark Matter}",
    booktitle = "{33rd International Cosmic Ray Conference}",
    eprint = "1307.7983",
    archivePrefix = "arXiv",
    primaryClass = "astro-ph.HE",
    pages = "0622",
    month = "7",
    year = "2013"
}

@article{Cohen:2016uyg,
    author = "Cohen, Timothy and Murase, Kohta and Rodd, Nicholas L. and Safdi, Benjamin R. and Soreq, Yotam",
    title = "{{\ensuremath{\gamma}} -ray Constraints on Decaying Dark Matter and Implications for IceCube}",
    eprint = "1612.05638",
    archivePrefix = "arXiv",
    primaryClass = "hep-ph",
    reportNumber = "MIT-CTP-4863, MIT-CTP 4863",
    doi = "10.1103/PhysRevLett.119.021102",
    journal = "Phys. Rev. Lett.",
    volume = "119",
    number = "2",
    pages = "021102",
    year = "2017"
}

@article{Fermi-LAT:2014ryh,
    author = "Ackermann, M. and others",
    collaboration = "Fermi-LAT",
    title = "{The spectrum of isotropic diffuse gamma-ray emission between 100 MeV and 820 GeV}",
    eprint = "1410.3696",
    archivePrefix = "arXiv",
    primaryClass = "astro-ph.HE",
    doi = "10.1088/0004-637X/799/1/86",
    journal = "Astrophys. J.",
    volume = "799",
    pages = "86",
    year = "2015"
}

@article{Fornasa:2015qua,
    author = "Fornasa, Mattia and S{\'a}nchez-Conde, Miguel A.",
    title = "{The nature of the Diffuse Gamma-Ray Background}",
    eprint = "1502.02866",
    archivePrefix = "arXiv",
    primaryClass = "astro-ph.CO",
    doi = "10.1016/j.physrep.2015.09.002",
    journal = "Phys. Rept.",
    volume = "598",
    pages = "1--58",
    year = "2015"
}

@article{Cirelli:2010xx,
    author = "Cirelli, Marco and Corcella, Gennaro and Hektor, Andi and Hutsi, Gert and Kadastik, Mario and Panci, Paolo and Raidal, Martti and Sala, Filippo and Strumia, Alessandro",
    title = "{PPPC 4 DM ID: A Poor Particle Physicist Cookbook for Dark Matter Indirect Detection}",
    eprint = "1012.4515",
    archivePrefix = "arXiv",
    primaryClass = "hep-ph",
    reportNumber = "CERN-PH-TH-2010-057, SACLAY-T10-025, IFUP-TH-2010-44",
    doi = "10.1088/1475-7516/2012/10/E01",
    journal = "JCAP",
    volume = "03",
    pages = "051",
    year = "2011",
    note = "[Erratum: JCAP 10, E01 (2012)]"
}

@article{Dominguez:2010bv,
    author = "Dominguez, A. and others",
    title = "{Extragalactic Background Light Inferred from AEGIS Galaxy SED-type Fractions}",
    eprint = "1007.1459",
    archivePrefix = "arXiv",
    primaryClass = "astro-ph.CO",
    doi = "10.1111/j.1365-2966.2010.17631.x",
    journal = "Mon. Not. Roy. Astron. Soc.",
    volume = "410",
    pages = "2556",
    year = "2011"
}

@article{Korsmeier:2022cwp,
    author = "Korsmeier, Michael and Pinetti, Elena and Negro, Michela and Regis, Marco and Fornengo, Nicolao",
    title = "{Flat-spectrum Radio Quasars and BL Lacs Dominate the Anisotropy of the Unresolved Gamma-Ray Background}",
    eprint = "2201.02634",
    archivePrefix = "arXiv",
    primaryClass = "astro-ph.HE",
    reportNumber = "FERMILAB-PUB-22-038-T",
    doi = "10.3847/1538-4357/ac6c85",
    journal = "Astrophys. J.",
    volume = "933",
    number = "2",
    pages = "221",
    year = "2022"
}

@article{Ajello:2020zna,
    author = "Ajello, M. and Di Mauro, M. and Paliya, V. S. and Garrappa, S.",
    title = "{The $\gamma$-Ray Emission of Star-forming Galaxies}",
    eprint = "2003.05493",
    archivePrefix = "arXiv",
    primaryClass = "astro-ph.GA",
    doi = "10.3847/1538-4357/ab86a6",
    journal = "Astrophys. J.",
    volume = "894",
    number = "2",
    pages = "88",
    year = "2020"
}

@article{Gruppioni:2013jna,
    author = "Gruppioni, C. and others",
    title = "{The Herschel PEP/HerMES Luminosity Function. I: Probing the Evolution of PACS selected Galaxies to z{\textasciitilde}4}",
    eprint = "1302.5209",
    archivePrefix = "arXiv",
    primaryClass = "astro-ph.CO",
    doi = "10.1093/mnras/stt308",
    journal = "Mon. Not. Roy. Astron. Soc.",
    volume = "432",
    pages = "23",
    year = "2013"
}

@article{Blanco:2021icw,
    author = "Blanco, Carlos and Linden, Tim",
    title = "{Star-forming galaxies provide a larger contribution to the isotropic gamma-ray background than misaligned active galactic nuclei}",
    eprint = "2104.03315",
    archivePrefix = "arXiv",
    primaryClass = "astro-ph.HE",
    doi = "10.1088/1475-7516/2023/02/003",
    journal = "JCAP",
    volume = "02",
    pages = "003",
    year = "2023"
}

@article{Linden:2016fdd,
    author = "Linden, Tim",
    title = "{Star-Forming Galaxies Significantly Contribute to the Isotropic Gamma-Ray Background}",
    eprint = "1612.03175",
    archivePrefix = "arXiv",
    primaryClass = "astro-ph.HE",
    doi = "10.1103/PhysRevD.96.083001",
    journal = "Phys. Rev. D",
    volume = "96",
    number = "8",
    pages = "083001",
    year = "2017"
}

@article{Koers:2008ba,
    author = "Koers, Hylke B. J. and Tinyakov, Peter",
    title = "{Testing large-scale (an)isotropy of ultra-high energy cosmic rays}",
    eprint = "0812.0860",
    archivePrefix = "arXiv",
    primaryClass = "astro-ph",
    reportNumber = "ULB-TH-08-38",
    doi = "10.1088/1475-7516/2009/04/003",
    journal = "JCAP",
    volume = "04",
    pages = "003",
    year = "2009"
}

@article{Siegal-Gaskins:2010oks,
    author = "Siegal-Gaskins, Jennifer M. and Reesman, Rebecca and Pavlidou, Vasiliki and Profumo, Stefano and Walker, Terry P.",
    title = "{Anisotropies in the gamma-ray sky from millisecond pulsars}",
    eprint = "1011.5501",
    archivePrefix = "arXiv",
    primaryClass = "astro-ph.HE",
    doi = "10.1111/j.1365-2966.2011.18672.x",
    journal = "Mon. Not. Roy. Astron. Soc.",
    volume = "415",
    pages = "1074--1082",
    year = "2011"
}

@article{Bierlich:2022pfr,
    author = "Bierlich, Christian and others",
    title = "{A comprehensive guide to the physics and usage of PYTHIA 8.3}",
    eprint = "2203.11601",
    archivePrefix = "arXiv",
    primaryClass = "hep-ph",
    reportNumber = "LU-TP 22-16, MCNET-22-04, FERMILAB-PUB-22-227-SCD",
    doi = "10.21468/SciPostPhysCodeb.8",
    journal = "SciPost Phys. Codeb.",
    volume = "2022",
    pages = "8",
    year = "2022"
}

@article{Cirelli:2009vg,
    author = "Cirelli, Marco and Panci, Paolo",
    title = "{Inverse Compton constraints on the Dark Matter e+e- excesses}",
    eprint = "0904.3830",
    archivePrefix = "arXiv",
    primaryClass = "astro-ph.CO",
    reportNumber = "SACLAY-T09-030",
    doi = "10.1016/j.nuclphysb.2009.06.034",
    journal = "Nucl. Phys. B",
    volume = "821",
    pages = "399--416",
    year = "2009"
}

@article{Meade:2009iu,
    author = "Meade, Patrick and Papucci, Michele and Strumia, Alessandro and Volansky, Tomer",
    title = "{Dark Matter Interpretations of the e+- Excesses after FERMI}",
    eprint = "0905.0480",
    archivePrefix = "arXiv",
    primaryClass = "hep-ph",
    reportNumber = "IFUP-TH-2009-9",
    doi = "10.1016/j.nuclphysb.2010.01.012",
    journal = "Nucl. Phys. B",
    volume = "831",
    pages = "178--203",
    year = "2010"
}

@article{Fornal:2020poq,
    author = "Fornal, Bartosz and Hewitt, Alec and Zhao, Yue",
    title = "{Baryonic and Leptonic GeV Dark Matter}",
    eprint = "2011.09014",
    archivePrefix = "arXiv",
    primaryClass = "hep-ph",
    doi = "10.1016/j.physletb.2021.136151",
    journal = "Phys. Lett. B",
    volume = "815",
    pages = "136151",
    year = "2021"
}

@article{Farina:2016ndq,
    author = "Farina, Marco and Monteux, Angelo and Shin, Chang Sub",
    title = "{Twin mechanism for baryon and dark matter asymmetries}",
    eprint = "1604.08211",
    archivePrefix = "arXiv",
    primaryClass = "hep-ph",
    doi = "10.1103/PhysRevD.94.035017",
    journal = "Phys. Rev. D",
    volume = "94",
    number = "3",
    pages = "035017",
    year = "2016"
}

@article{Kilic:2021zqu,
    author = "Kilic, Can and Verhaaren, Christopher B. and Youn, Taewook",
    title = "{Twin quark dark matter from cogenesis}",
    eprint = "2109.03248",
    archivePrefix = "arXiv",
    primaryClass = "hep-ph",
    reportNumber = "UTTG 16-2021",
    doi = "10.1103/PhysRevD.104.116018",
    journal = "Phys. Rev. D",
    volume = "104",
    number = "11",
    pages = "116018",
    year = "2021"
}

@article{Zhao:2014nsa,
    author = "Zhao, Yue and Zurek, Kathryn M.",
    title = "{Indirect Detection Signatures for the Origin of Asymmetric Dark Matter}",
    eprint = "1401.7664",
    archivePrefix = "arXiv",
    primaryClass = "hep-ph",
    reportNumber = "MCTP-14-02",
    doi = "10.1007/JHEP07(2014)017",
    journal = "JHEP",
    volume = "07",
    pages = "017",
    year = "2014"
}

@article{Davoudiasl:2010am,
    author = "Davoudiasl, Hooman and Morrissey, David E. and Sigurdson, Kris and Tulin, Sean",
    title = "{Hylogenesis: A Unified Origin for Baryonic Visible Matter and Antibaryonic Dark Matter}",
    eprint = "1008.2399",
    archivePrefix = "arXiv",
    primaryClass = "hep-ph",
    doi = "10.1103/PhysRevLett.105.211304",
    journal = "Phys. Rev. Lett.",
    volume = "105",
    pages = "211304",
    year = "2010"
}

@article{Alonso-Alvarez:2021oaj,
    author = "Alonso-{\'A}lvarez, Gonzalo and Elor, Gilly and Escudero, Miguel and Fornal, Bartosz and Grinstein, Benjam{\'\i}n and Martin Camalich, Jorge",
    title = "{Strange physics of dark baryons}",
    eprint = "2111.12712",
    archivePrefix = "arXiv",
    primaryClass = "hep-ph",
    reportNumber = "TUM-HEP-1373/21, MITP-21-060",
    doi = "10.1103/PhysRevD.105.115005",
    journal = "Phys. Rev. D",
    volume = "105",
    number = "11",
    pages = "115005",
    year = "2022"
}

@article{Cirelli:2005uq,
    author = "Cirelli, Marco and Fornengo, Nicolao and Strumia, Alessandro",
    title = "{Minimal dark matter}",
    eprint = "hep-ph/0512090",
    archivePrefix = "arXiv",
    reportNumber = "DFTT40-2005, IFUP-TH-2005-34",
    doi = "10.1016/j.nuclphysb.2006.07.012",
    journal = "Nucl. Phys. B",
    volume = "753",
    pages = "178--194",
    year = "2006"
}

@article{Cirelli:2007xd,
    author = "Cirelli, Marco and Strumia, Alessandro and Tamburini, Matteo",
    title = "{Cosmology and Astrophysics of Minimal Dark Matter}",
    eprint = "0706.4071",
    archivePrefix = "arXiv",
    primaryClass = "hep-ph",
    reportNumber = "IFUP-TH-2007-12, SACLAY-T07-052",
    doi = "10.1016/j.nuclphysb.2007.07.023",
    journal = "Nucl. Phys. B",
    volume = "787",
    pages = "152--175",
    year = "2007"
}

@article{Cirelli:2009uv,
    author = "Cirelli, Marco and Strumia, Alessandro",
    title = "{Minimal Dark Matter: Model and results}",
    eprint = "0903.3381",
    archivePrefix = "arXiv",
    primaryClass = "hep-ph",
    reportNumber = "IFUP-TH-2009-04, SACLAY-T09-010",
    doi = "10.1088/1367-2630/11/10/105005",
    journal = "New J. Phys.",
    volume = "11",
    pages = "105005",
    year = "2009"
}

@article{Garcia-Cely:2015dda,
    author = "Garcia-Cely, Camilo and Ibarra, Alejandro and Lamperstorfer, Anna S. and Tytgat, Michel H. G.",
    title = "{Gamma-rays from Heavy Minimal Dark Matter}",
    eprint = "1507.05536",
    archivePrefix = "arXiv",
    primaryClass = "hep-ph",
    doi = "10.1088/1475-7516/2015/10/058",
    journal = "JCAP",
    volume = "10",
    pages = "058",
    year = "2015"
}

@article{Fan:2013faa,
    author = "Fan, JiJi and Reece, Matthew",
    title = "{In Wino Veritas? Indirect Searches Shed Light on Neutralino Dark Matter}",
    eprint = "1307.4400",
    archivePrefix = "arXiv",
    primaryClass = "hep-ph",
    doi = "10.1007/JHEP10(2013)124",
    journal = "JHEP",
    volume = "10",
    pages = "124",
    year = "2013"
}

@article{Cirelli:2015bda,
    author = "Cirelli, Marco and Hambye, Thomas and Panci, Paolo and Sala, Filippo and Taoso, Marco",
    title = "{Gamma ray tests of Minimal Dark Matter}",
    eprint = "1507.05519",
    archivePrefix = "arXiv",
    primaryClass = "hep-ph",
    reportNumber = "SACLAY-T15-130, ULB-PHYS-TH-15-16",
    doi = "10.1088/1475-7516/2015/10/026",
    journal = "JCAP",
    volume = "10",
    pages = "026",
    year = "2015"
}

@article{Garny:2012vt,
    author = "Garny, Mathias and Ibarra, Alejandro and Tran, David",
    title = "{Constraints on Hadronically Decaying Dark Matter}",
    eprint = "1205.6783",
    archivePrefix = "arXiv",
    primaryClass = "hep-ph",
    reportNumber = "TUM-HEP-840-12, DESY-12-084, FTPI-MINN-12-18, UMN-TH-3104-12",
    doi = "10.1088/1475-7516/2012/08/025",
    journal = "JCAP",
    volume = "08",
    pages = "025",
    year = "2012"
}

@article{Ibarra:2013cra,
    author = "Ibarra, Alejandro and Tran, David and Weniger, Christoph",
    title = "{Indirect Searches for Decaying Dark Matter}",
    eprint = "1307.6434",
    archivePrefix = "arXiv",
    primaryClass = "hep-ph",
    doi = "10.1142/S0217751X13300408",
    journal = "Int. J. Mod. Phys. A",
    volume = "28",
    pages = "1330040",
    year = "2013"
}

@article{Cheng:2016slx,
    author = "Cheng, Hsin-Chia and Huang, Wei-Chih and Huang, Xiaoyuan and Low, Ian and Tsai, Yue-Lin Sming and Yuan, Qiang",
    title = "{AMS-02 Positron Excess and Indirect Detection of Three-body Decaying Dark Matter}",
    eprint = "1608.06382",
    archivePrefix = "arXiv",
    primaryClass = "hep-ph",
    doi = "10.1088/1475-7516/2017/03/041",
    journal = "JCAP",
    volume = "03",
    pages = "041",
    year = "2017"
}

@article{Dreiner:2008tw,
    author = "Dreiner, Herbi K. and Haber, Howard E. and Martin, Stephen P.",
    title = "{Two-component spinor techniques and Feynman rules for quantum field theory and supersymmetry}",
    eprint = "0812.1594",
    archivePrefix = "arXiv",
    primaryClass = "hep-ph",
    reportNumber = "BN-TH-2008-12, SCIPP-08-08, FERMILAB-PUB-09-855-T, BN-TH-2008-12 and SCIPP-08/08",
    doi = "10.1016/j.physrep.2010.05.002",
    journal = "Phys. Rept.",
    volume = "494",
    pages = "1--196",
    year = "2010"
}

@article{Doroshkevich:1984,
    author = "Doroshkevich, A. G. and Khlopov, M. Yu.",
    title = "{Formation of structure in a universe with unstable neutrinos}",
    journal = "Mon. Not. Roy. Astron. Soc.",
    volume = "211",
    pages = "277--282",
    year = "1984",
    doi = "10.1093/mnras/211.2.277"
}
\end{document}